\documentclass[preprint2]{aastex}
\def\chandra{\textsl{Chandra}}

\def\sun{\hbox{$\odot$}}
\def\farcm{\hbox{$.\mkern-4mu^\prime$}}
\def\farcs{\hbox{$.\!\!^{\prime\prime}$}}

\slugcomment{Submitted to the Astronomical Journal.}

\shorttitle{ChAInGeS: The Chandra Arp Interacting Galaxies Survey}
\shortauthors{Smith et al.}

\begin{document}


\title{ChAInGeS: The Chandra Arp Interacting Galaxies Survey}
\shortauthors{Smith }


\author{Beverly J. Smith\altaffilmark{1},
Douglas A. Swartz\altaffilmark{2},
Olivia Miller\altaffilmark{1},
Jacob A. Burleson\altaffilmark{3},
Michael A. Nowak\altaffilmark{4}, and
Curtis Struck\altaffilmark{5}}

\altaffiltext{1}{Department of 
Physics and Astronomy, East Tennessee
State University, Johnson City TN  37614;
smithbj@etsu.edu; millero@goldmail.etsu.edu}
\altaffiltext{2}{University Space Research
Association, NASA Marshall Space Flight Center, VP 62, Huntsville AL,
Douglas.A.Swartz@nasa.gov}
\altaffiltext{3}{University of Alabama Huntsville AL  35805; jab0039@uah.edu}
\altaffiltext{4}{MIT-CXC, mnowak@space.mit.edu}
\altaffiltext{5}{Department of Physics and Astronomy, Iowa State University, Ames IA  50011; curt@iastate.edu}



\begin{abstract}

We have conducted a statistical 
analysis of the ultra-luminous X-ray point sources (ULXs; L$_X$ $\ge$ 
10$^{39}$ erg/s) in
a sample of galaxies selected from the Arp Atlas of Peculiar Galaxies.
We find a possible enhancement of a factor of $\sim$2$-$4 in 
the number of ULXs per blue luminosity 
for the strongly interacting subset.
Such an enhancement would be expected if ULX production
is related to star formation, as interacting galaxies
tend to have enhanced star formation rates on average.
For most of the Arp galaxies in our sample,
the total number of ULXs compared to the far-infrared luminosity
is consistent with values found earlier for spiral galaxies.
This suggests that for these galaxies, ULXs trace recent star formation.
However, for the most infrared-luminous
galaxies,
we find a deficiency of ULXs 
compared to the infrared luminosity.   For these very infrared-luminous
galaxies, AGNs
may contribute to powering the far-infrared; alternatively, 
ULXs
may be highly obscured in the X-ray
in these galaxies
and therefore not detected by these Chandra observations.
We determined
local
UV/optical colors within the galaxies in the 
vicinity of the candidate ULXs using GALEX UV and SDSS optical images.
In most cases, the distributions of colors are similar to 
the global colors of
interacting galaxies.
However, the u $-$ g and r $-$ i colors at the ULX locations tend to be 
bluer on average than these global colors, 
suggesting that ULXs are preferentially found in regions
with young stellar
populations.
In the Arp sample there is a possible enhancement of a factor of $\sim$2 $-$ 5 in 
the fraction of galactic nuclei that are X-ray bright 
compared to more normal spirals.

\end{abstract}


%% subject header.  Objects should be in the appropriate "individual"
%% headers (e.g. quasars: individual, stars: individual, etc.) with the
%% additional provision that the total number of headers, including each
%% individual object, not exceed six.  The \objectname{} macro, and its
%% alias \object{}, is used to mark each object.  The macro takes the object
%% name as its primary argument.  This name will appear in the paper
%% and serve as the link's anchor in the electronic edition if the name
%% is recognized by the data centers.  The macro also takes an optional
%% argument in parentheses in cases where the data center identification
%% differs from what is to be printed in the paper.


\section{Introduction}

In the dozen years since its launch, 
the \chandra\ X-ray Observatory has discovered
hundreds of ultra-luminous ($\ge$10$^{39}$ erg/s) X-ray point sources
(ULXs) in external galaxies.
Several different scenarios have been suggested
to explain these sources, 
including sub-Eddington
intermediate mass (100 - 1000 M$_{\sun}$) black holes 
\citep{colbert99, portegies04a, portegies04b, freitag06}
or stellar mass black holes 
with beamed
\citep{king01, king09}
or super-Eddington \citep{begelman02} X-ray emission.
In a few cases
sensitive high resolution optical images
have found late O or early B stars that are the apparent mass donors
for the ULXs, suggesting stellar mass black holes
\citep{liu02, liu04, kuntz05, ptak06, terashima06,
roberts08, grise08, tao11}.  In other cases, emission nebulae have been found
that are apparently associated with the ULX \citep{pakull03, pakull06}.
However, in most cases a discrete optical counterpart to the 
ULX is not found.

Statistical studies of the environments of ULXs provide additional
clues to their nature.
The number of ULXs in spiral galaxies has been found to be correlated
with the galaxy's global star formation rate, suggesting
that they are mostly high mass X-ray binaries (HMXBs) \citep{swartz04, swartz11,
liu06}.
This is consistent with studies that show good correlations of
the star formation rate of a galaxy with
its global hard X-ray luminosity 
\citep{ranalli03, persic07} and with the number of HMXBs \citep{grimm03, mineo12}.
For ellipticals and low mass spirals, the ULX frequency seems
to be better correlated with the stellar mass, and the ULX luminosities
tend to be lower, suggesting that
these sources may be the high luminosity end of the low mass X-ray
binary population (LMXBs) \citep{swartz04}.
In general, LMXBs appear to be good tracers of the stellar
mass of a galaxy \citep{gilfanov04}.
\citet{swartz08} found that ULXs tend to be relatively abundant
in dwarf galaxies, suggesting that the ULX frequency per stellar
mass decreases with increasing galaxian mass.
In a sample of very nearby galaxies,
\citet{swartz09} found that the local environment (100 pc $\times$ 100 pc) 
around
the ULX tended to have bluer optical colors on average, indicating
a link with young stellar populations.
Based on these colors, they
suggest that the most luminous ULXs are associated with 
early B-type stars with ages of $\approx$10 $-$ 20 Myrs.
They argue that most ULXs
are stellar mass black holes rather than more massive
objects. 

A recent survey of nearby galaxies
suggests that there is a cut-off in the
ULX X-ray luminosity function at high luminosities, with
few sources above 2 $\times$ 10$^{40}$ erg/s
\citep{swartz11}.   However, 
two recent studies have identified some extremely luminous
($\ge$ 10$^{41}$ erg/s)
X-ray point sources in more distant 
galaxies \citep{farrell09, sutton11}.
Such extreme luminosities can best be explained by intermediate
mass black holes,
as black holes with $\le$80 M$_{\sun}$ are not expected to produce 
such high luminosities \citep{swartz11}.
Extending the current surveys of nearby galaxies
to more distant and more luminous galaxies is critical to better constrain
the upper end of this luminosity function.

There is some evidence that gravitational interactions between
galaxies may increase the numbers of ULXs present in a galaxy.
Studies of individual interacting and merging galaxies show
that some have 
numerous ULXs (e.g.,
Arp 244,
\citealp{zezas02}; Arp 284, \citealp{smith05};
NGC 3256, \citealp{lira02}; Arp 269, \citealp{roberts02};
Arp 270, \citealp{brassington05};
NGC 4410, \citealp{smith03}; the Cartwheel, \citealp{gao03, wolter06,
crivellari09}).
Since interacting
galaxies tend to have enhanced star formation on average compared
to normal galaxies (e.g., 
\citealp{bushouse87, bushouse88,
kennicutt87, barton00, barton03, smith07, ellison08}), one might expect
enhanced ULX production in interacting systems
if ULXs are associated with star formation.

Other conditions
within interacting and merging galaxies may also increase the number
of ULXs.
For example,
\citet{soria06}
have suggested that cloud collisions during
galaxy interactions may lead to rapid, large-scale collapse of 
molecular clouds, producing a few massive protostars which 
quickly coalesce into a single star with mass $>$ 100 M$_{\sun}$.
This process may lead to an excess of ULXs in interacting systems.
If intermediate mass black holes form in dense stellar clusters
\citep{portegies04a}, and such clusters are more likely formed
in interacting and merging galaxies due to increased gas pressure
and enhanced cloud collisions, 
then ULXs might be expected to be found
more commonly in interacting systems
if such black holes contribute 
significantly
to the ULX population.
We note that in Arp 244 four ULX candidates are
seen in the `overlap' region where 
the two disks intersect \citep{zezas02},
a region where strong shocks and/or compression may
be present.
Alternatively, black hole fueling rates may be higher 
in interacting galaxies due to the more disturbed orbits of 
the interstellar clouds.
Thus there are a number of theoretical ideas for why ULXs might
be more common in interacting and merging galaxies.

In some interacting systems, X-ray sources have been
found that are apparently associated
with tidal features \citep{smith05, brassington05}.
If there is a statistically-significant correlation between
ULXs and tidal features, it would imply that 
the environment of tidal features is particularly 
inducive to the formation of ULXs.
On average, tidal features tend to have somewhat bluer UV/optical
colors than their parent galaxies \citep{smith10}, possibly
due to higher mass-normalized star formation.
Strong gas compression and shocking can occur
along tidal features
(e.g., 
\citealp{gerber94, struck03}),
possibly inducing star formation and ULX creation.

In the current study,
we present results from an archival \chandra\ survey
of a large sample of interacting galaxies,
comparing with 
available optical, UV, and infrared data.
We use this database to
investigate whether ULXs are more likely to occur in 
interacting 
galaxies than in normal systems, relative to the stellar mass and
star formation rate, and 
whether they are more likely to occur
in tidal features compared to disks.
We also
determined local UV/optical colors 
within the galaxies in the vicinity of the ULX to investigate
the local stellar population. 
In addition, we use this dataset
to determine the frequency of X-ray bright nuclei in
interacting galaxies.

In Section 2 of this paper, we describe the galaxy sample, the Chandra
datasets, and the ancilliary UV, optical, and infrared data.
In Section 3, we describe the ULX sample.  
In 
Section 4, we compare the number of ULXs with optical luminosity
as a proxy for stellar mass, and in Section 5, we compare with 
previous studies of mostly spiral galaxies.
In Section 6, we define a subset of strongly interacting galaxies
and investigate their ULX properties.   In Section 7 we compare
with the far-infrared luminosity.  
We compare
UV/optical colors in the vicinity of the ULX candidates to global
colors of galaxies in Section 8.
Finally, 
we obtain statistics on the frequency
of X-ray detected nuclei in this sample compared to spiral samples
in Section 9.

\section{Sample Selection and Additional Data}

To obtain a large enough sample of interacting galaxies for a statistical
analysis of their ULXs, 
we started with the 338 systems in 
the \citet{arp66} Atlas of Peculiar Galaxies.  
This Atlas contains most of the strongly interacting galaxies in the
local Universe
that are close enough for a detailed spatially-resolved analysis.
Searching the \chandra\ archives,
we found that 95 Arp Atlas
systems have archival ACIS-S or ACIS-I Chandra observations
pointed at a location within 10$'$ of the 
SIMBAD\footnote{http://simbad.u-strasbg.fr}/NASA Extragalactic Database
(NED\footnote{http://nedwww.ipac.caltech.edu}) coordinates 
of the system.
We eliminated systems
with sensitivities poorer than 
10$^{40}$
erg/s 
in the 
0.5 $-$ 8 keV range
at the distance of the host galaxy, assuming a 10-count limit.
For observations with the ACIS-S array, we only used datasets in
which the S3 chip covered 
at least part of the Arp system.
For ACIS-I observations, at least one of the four I array chips
must have covered at least part of the Arp system for the system
to be included in our sample.
This ensures uniform sensitivity by omitting galaxies
that are far off-axis.
The aimpoint for the ACIS-S array is located in the S3 chip,
so it has the highest sensitivity,
while
the aimpoint for the ACIS-I array is close to the 
center of the array,
so no one chip is strongly favored in sensitivity.

We searched the 
Sloan Digitized Sky Survey (SDSS; \citealp{abazajian03})
archives for optical images of these
galaxies.   In our final sample, we only included galaxies
which had calibrated SDSS Data Release 7 (DR7) images available.
In some cases, the Arp system is split between
two or three SDSS images.    
In these cases, we treated each SDSS image separately.

There were 45 Arp systems that fit these criteria.
These systems are listed in Table 1, along with their positions,
distances, and angular sizes.  
The distances were obtained
from 
NED, using, as a first preference,
the mean of the redshift-independent determinations, and as a second
choice, H$_0$ = 73 km/s/Mpc, with Virgo,
Great Attractor, and SA infall models.
The angular size given
is the total angular extent of the Arp system, obtained
from NED, when available. 
Otherwise, it was estimated from the SDSS images or the 
Digital Sky Survey images
available from NED.
Most of these galaxies are relatively nearby,
so \chandra\ provides
good spatial information within the galaxies, yet most also
have relatively small angular separations, so 
fit within the \chandra\ S3 8\farcm3 $\times$ 8\farcm3 field of view.  

There are a total of 69 individual galaxies in the 45 Arp systems,
that are covered at least in part by the Chandra S3 or I array
field of view.  
The names of the individual galaxies observed
by Chandra are provided in Table 1.
Of these 69 galaxies,
49 (69\%) are classified as spirals in NED,
9 (13\%) are irregular, and 8 (12\%) are elliptical, with the rest peculiar
or not typed.
For most of the merger remnants in our sample (Arp 155,
160, 193, 215, 217, and 243), only a single nucleus
is visible in near-infrared images \citep{rothberg04, haan11}
and in high spatial resolution radio continuum maps
\citep{lucero07, parma86, krips07, duric86, parra10},
thus we consider them a single galaxy for this study.
For Arp 220, two nuclei are visible
in near-infrared images \citep{graham90}
and radio images \citep{norris88}, thus we consider it
a galaxy pair.
Arp 299 has three peaks in K band \citep{bushouse92} and in the 
mid-infrared \citep{gehrz83}, 
all with counterparts in the radio 
continuum \citep{condon82, gehrz83}.
However, follow-up multi-wavelength observations
suggest that the third source may be an extranuclear star forming
region rather than the nucleus of a third galaxy 
\citep{dudley93, imanishi06, alonso09}.
Thus we consider Arp 299 two galaxies rather
than three.  This information is used in
Section 9 of this paper, when we determine the fraction of galactic
nuclei in this sample that are detected in the X-ray.
For Arp 189, the Chandra S3 field of view covers
the tidal tail but not the main body of the galaxy.
Four of the 69 galaxies are 
listed as Seyfert 1 or 2 in
NED, 20 are listed as LINERs or transition objects (i.e., Sy/LINER
or LINER/HII),
and 17 are listed as having HII-type nuclear spectra.
The nuclear spectral types are given in Table 1. 

Table 1 also lists
the \chandra\ exposure time and the corresponding
X-ray point source luminosity
(0.5 $-$ 8 keV) detection limit assuming
a 10-count limit.  This limit was calculated using the Portable
Interacting Multi-Mission Simulator (PIMMS; \citealp{mukai93}) assuming
an absorbed power law with spectral index $\Gamma$ = 1.7 and Galactic
HI column densities from \citet{schlegel98}.
In some cases, more than one \chandra\
observation is available for a given system.   
We generally used the dataset best-centered on the system,
or the longest exposure.

We also searched the Galaxy Evolution
Explorer (GALEX) archives
for UV images of these galaxies,
selecting only GALEX images with exposure
times greater than 1000 sec.   
Of our 45 systems, 28 have appropriate GALEX images available.
These systems are listed in Table 1, along with their GALEX exposure time.
The
GALEX NUV and FUV FWHM values 
are 5\farcs6 and 4\farcs0, respectively, and 
the GALEX pixels
are 1\farcs5 across.   
The field of view of the GALEX images is about 1\fdg2 across.
For comparison,
the 
SDSS images have full width half maximum (FWHM) spatial resolutions
between 0\farcs8 and 2\farcs2, pixel sizes of 0\farcs4, and
a field of view of 13\farcm6 $\times$
9\farcm9.

We also obtained 
Infrared Astronomical Satellite (IRAS)
total 60 and 100 $\mu$m flux densities for these systems
from \citet{rice88}, \citet{surace04}, \citet{xu03},
or the xscanpi 
program\footnote{http://scanpiops.ipac.caltech.edu:9000/applications/Scanpi/},
being careful to include the total flux from the galaxy or galaxies covered
by the Chandra field of view.
The far-infrared luminosity of the included galaxy or galaxies for
each Arp system is also provided in Table 1.  These values
were calculated using 
$L_{FIR} = 3.87 \times 10^{5}D^2(2.58F_{60} + F_{100}$),
where $F_{60}$ and $F_{100}$ are the 60 and 100 $\mu$m flux densities
in Jy, D is the distance in Mpc, and L$_{FIR}$ is in L$_{\sun}$.
For Stephan's Quintet (Arp 319), we excluded the flux from NGC 7320,
which is a foreground galaxy.
These luminosities are utilized in Section 7.

\section{ULX Catalog}

\subsection{X-Ray Point Source Identification}

Because of the wide range in X-ray sensitivities across our sample,
 individual X-ray sources may have few detected counts even though
 their estimated luminosities may be high.
For this reason, we used three methods of source detection.
First, we followed the data reduction procedure of Smith et al. (2005)
 using CIAO version 4.2 and the
latest calibration files, additionally removing pixel randomization
 in creating the Level 2 event lists.
On these datasets, we
then used both the CIAO utility {\sl wavdetect}
and the source finding algorithm described in Tennant (2006) to find
X-ray point sources.   In addition, we used
the Chandra Source Catalog (CSC; \citealp{evans10})
to search for X-ray point sources in these galaxies.
In a few cases, sources listed in the CSC were not found as point sources
by the source extraction routines listed above.
These sources were generally listed as extended in the CSC,
and appeared diffuse in the Level 2 event images, thus are not classified
as ULXs.
In addition,
some sources were not included 
in the CSC because they lay on CCD chips that contained high levels
of extended diffuse X-ray emission\footnote{see 
http://cxc.harvard.edu/csc/faq/dropped$\_$chips.txt}.
Others were not included in the CSC because the observations
only became publicly available recently or were made in the 
spatially-restricted sub-array mode. 

For the following analysis,
we used the CSC determination of the ACIS aperture-corrected 
broadband (0.5 - 7 keV) flux, 
when available,
after
converting these values to 
0.5 $-$ 8.0 keV assuming a power law with photon index $\Gamma$ = 1.7.
Otherwise, the same procedure as was used to estimate detection limits
was applied to the net
observed source counts obtained from the \citet{tennant06}
source-finding results.
In this work, we use 
the traditional definition of 
a ULX, with a 0.5 $-$ 8 keV 
X-ray luminosity $\ge$
10$^{39}$ erg/s 
(e.g., \citealp{swartz04}).
We corrected these X-ray fluxes for Galactic extinction using
the \citet{schlegel98} Galactic HI column densities.
Together, these two corrections are 
typically small, a $\sim$10\% effect. 
We did not correct for internal extinction.
In most cases, the available archival
X-ray data was not sensitive enough for a detailed spectroscopic
analysis and derivation of the internal extinction. 
As noted by \citet{swartz04} and \citet{swartz11}, 
correction for such internal extinction may increase
the derived L$_X$ values significantly for some objects.

\subsection{Classification of the Candidate ULXs}

As a first step in our analysis, we classified our
X-ray sources as `tidal', `disk', `nuclear' or `off-galaxy' depending on their
location relative to optical features.
We used a cut-off
of 2.5 counts above sky 
on the smoothed, sky-subtracted SDSS g images
as our dividing line between `off-galaxy' and `galaxy'.
This corresponds to approximately 25.7 magnitudes/sq.\ arcsec, but this
varies slightly from image to image.

For the systems
observed with the ACIS-S array,
we only included sources 
that lie within the Chandra ACIS-S S3 chip field of view.
In some datasets, other
chips were read out as well, however, for consistency and for 
ease in determining the observed sky coverage needed to
calculate background contamination, we only used
sources in the S3 field of view.
For the ACIS-I observations, we used sources in all four I-array chips.

The distinction between `disk' and `tidal' sources is sometimes ambiguous,
as our determination of where the tail/bridge begins is subjective.
This is especially true in the case of spiral arms, which may
be tidally-disturbed by the interaction.  
In ambiguous cases, 
when the 
outer contours of the smoothed g band
image showed a smoothly elliptical shape, we classified it as 
a spiral arm within a disk.  In contrast, if 
considerable asymmetry was seen in the outermost contours, we classified
the structure as tidal.
As in \citet{smith07},
we used multiple rectangular boxes to mark the extent of the 
tidal features and the disks.   We used these boxes to decide
whether a given source will be classified as tidal rather than disk. 

X-ray point sources were identified as `nuclear' if they were 
coincident
within 6 SDSS pixels ($\le$2\farcs4)
with the SDSS z-band peak 
in the inner disk of the galaxy.   
This is a conservative limit, as the 99\% uncertainty in the Chandra
absolute position for sources within 3$'$ of the aimpoint is 
quoted as 0\farcs8\footnote{http://asc.harvard.edu/cal/ASPECT/celmon/}, 
while the astrometric accuracy of the SDSS
images has been quoted as less than 1 pixel (0\farcs4) \citep{stoughton02}.
In most cases, the X-ray/optical alignment of
the nuclear sources is within 1$''$, but 
for Arp 259 and 318, the offsets are 2\farcs4.
In these cases, it is possible that the sources may be truly 
non-nuclear, thus we may be inadvertently eliminating
a few ULXs from the disk sample.

As an example
to illustrate our classification scheme,
in Figure 1 we display the smoothed SDSS g image of one of the Arp
systems in our sample, Arp 259.
The nuclear X-ray source
is marked as a green circle, and the tidal ULX candidate is a yellow circle.
The rectangular regions selected as disk are outlined in green,
while the disk areas included are marked in red.   The white contours
mark 2.5 counts above the sky level, our dividing
line between `sky' and `galaxy'.
According to \citet{amram07}, this system consists of three 
interacting galaxies.  
The tidal X-ray source is $\sim$3$''$ from an optical knot.

\subsection{X-Ray Source Sensitivity}

As can be seen in Table 1, for 20 of the 45 Arp systems in the sample 
the sensitivity of the data is above the ULX luminosity cut-off of
10$^{39}$ erg/s.
Because there is a range in the sensitivities of the archival
observations we are using, 
in the following analysis we divide the sample of galaxies into
two subsets which we analyze separately.
In the following discussion we refer to the original set of 45
Arp systems as the `intermediate sample', 
while the subset of 25 systems with
sensitivities $\le$10$^{39}$ erg/s we call the `sensitive sample'.
We will use the sensitive sample to provide statistics on the total
number of ULXs present, while the intermediate sample
will be used to provide constraints on the top of the X-ray luminosity function
($\ge$10$^{40}$ erg/s).
From the X-ray luminosity functions presented in
the literature for large samples of luminous
extragalactic X-ray sources,
a break or cut-off is
indicated at about 1 $-$ 2 $\times$ 10$^{40}$ erg/s 
\citep{grimm03, swartz04, swartz11,
liu11, mineo12}.
Thus 10$^{39}$ erg/s is the `complete' sample accounting for
all the ULXs in the sample galaxies, while $\ge$10$^{40}$ erg/s 
is the extreme luminous end of the `normal' ULX population up
to and above the luminosity function cutoff and may include
extremely luminous examples similar to the hyperluminous 
object reported by \citet{farrell09}.

In Table 2, we list the positions and 0.5 $-$ 8 keV X-ray 
luminosities of the
disk and tidal ULX candidates and the nuclear X-ray sources
identified in this study.
The off-galaxy X-ray sources are not included.
These sources are assigned to the sensitive sample and/or the intermediate
sample, depending upon which galaxy the sources are in,
and the sensitivity of the Chandra observation of that galaxy.
These assignments are provided in Table 2.
Note that a source
can belong to more than one sample, depending upon 
its luminosity and the sensitivity of the Chandra dataset.
In Table 2 and in the subsequent analysis, 
we do not include X-ray sources with 10$^{39}$ erg/s $\le$ L$_X$ $<$ 10$^{40}$ erg/s
for galaxies that are not in the `sensitive sample'
(i.e., we do not include sources in this luminosity range for galaxies with 
sensitivities $>$ 10$^{39}$ erg/s).

In Table 3, we list the numbers of Arp systems in both
the sensitive and intermediate
samples, as well as the total number of individual galaxies in those
systems that are included in the Chandra S3 or I array field of view.
Table 3 also gives the 
mean distance to the Arp systems in the sample and the limiting L$_X$
used for the sample.
Various additional properties of the two sets of galaxies are also
included in Table 3, and
are discussed below.

In Table 3, we list the total number of disk, tidal, nuclear, and off-galaxy
Chandra sources 
in each sample of galaxies, which have X-ray luminosities
above the cut-off luminosity for that sample.
In total, adding up over the two samples, and accounting
for the fact that some ULXs can be in both the sensitive and intermediate
samples,
we have a total of 
58 disk ULX candidates, 13 tidal sources, 26 nuclear
sources, and 64 off-galaxy objects.
No non-nuclear source in our sample of peculiar
galaxies has L$_X$ $>$ 10$^{41}$ erg/s,
thus we do not find any ULX candidates
with the extreme luminosities
of those found by \citet{farrell09} and \citet{sutton11}.

\subsection{Optical Counterparts to X-ray Sources}

We inspected the SDSS g images for optical counterparts to these
X-ray sources.  
The disk and tidal ULX candidates with optical counterparts 
may be background objects such as quasars, active nuclei, 
or small angular size galaxies, or foreground stars.
Alternatively, they may be ULXs associated with or located
near compact knots of star formation or star clusters 
within the Arp galaxy.  
Sources with discrete optical counterparts
are identified in Table 2, while
the number of X-ray sources in each category that
have optical counterparts is provided in Table 3.
As shown in Table 3,
the fraction of X-ray sources in the tails and disks that have
optical counterparts are 0\% to 44\%, depending upon
the sample, while 
35\% $-$ 52\% of the off-galaxy X-ray sources have optical counterparts.

For the off-galaxy sources with optical counterparts, we estimated 
optical magnitudes from the SDSS images and calculated the ratio 
of the X-ray flux to that in the optical, f$_X$/f$_V$, as defined 
by \citet{maccacaro88}, converting to the 
0.3 $-$ 3.5 keV flux using PIMMS with photon index 1.7 and using
the optical magnitude conversions of \citet{jester05}.
Almost all have f$_X$/f$_V$ $\ge$ 0.3, consistent 
with them being background AGN, though a few
appear to be foreground stars, which are
expected to have lower f$_X$/f$_V$ ratios
(e.g., \citealp{maccacaro88, della04}). 

For the disk/tidal ULX candidates, followup
optical observations are needed to confirm whether apparent optical 
counterparts are indeed likely to be associated with the ULXs 
and to accurately determine galaxy-subtracted optical magnitudes. 

\subsection{Background Contamination }

To properly compare the disk vs.\ tail/bridge numbers, 
we also need to correct
for contamination by background sources.
For background subtraction, we need an estimate of the 
surface area subtended by the tails/bridges vs.\ the disks.
To this end, we counted the number of SDSS pixels within
our tail/bridge vs.\ disk rectangular boxes, which are above our
surface brightness cut-off level on the smoothed sky-subtracted
g images (2.5 counts), that were also covered by either
the Chandra 
S3 field
of view (for ACIS-S observations) or 
any of the I chips (for ACIS-I observations).  
To accomplish this, we constructed a first mask of the sky, delineating the
location of our rectangular disk/tidal boxes, a second mask,
marking the location of the galaxy above the surface brightness
cut-off, and a third mask, showing the field of view of the
Chandra chip(s) in question.   These three masks were then
combined to create a single mask from which the number of
relevant pixels was determined.  These values were then converted
into square arcminutes using the SDSS pixel size.
In Table 3, for both samples of 
galaxies, we give the total surface area covered by
the survey, in arcminutes.
As expected, the area covered by the sensitive survey is
less than that covered by the intermediate survey.

To estimate the amount of contamination of our ULX sample
due to background X-ray sources, for each galaxy in each sample
we calculated
the flux corresponding 
to the selected
L$_X$ cut-off for that sample.  
We then estimated 
the number of probable background sources masquerading as ULXs
for each galaxy by   
multiplying the observed area for each galaxy by 
the 
\citet{moretti03} determination of  
the number
of background X-ray sources per area 
as a function of flux. 
We then summed 
over all of the galaxies in the respective sample to calculate
the total expected number
of background contaminants for that survey.
These numbers are given in Table 3, along with the estimated
fraction of the observed sources which are likely background objects.
These percentages range from $\sim$20\% for the disk and 
tidal sources
in the sensitive sample, to 31\% $-$ 40\% for the intermediate
sample.
In Table 3, we also provide the background-corrected ULX counts in
the disks and tails, after subtracting the expected number of interlopers.

\section{Number of ULXs per SDSS Luminosity }

Next, we scaled
the background-corrected numbers of ULX sources
by optical luminosity
as a proxy for stellar mass, to determine whether ULXs are preferentially
found in tidal regions or in inner disks.  
For this estimate, we used the SDSS r band
flux.
For a more direct comparison with the \citet{swartz04}
study, which used B luminosities, we also calculated the SDSS g flux.
We first sky-subtracted the SDSS images using the mean value from
several rectangular
sky regions far from the galaxies.  
We then determined the
total SDSS r and g band fluxes within our defined `disk' and `tidal' rectangular
boxes, including only pixels covered by the Chandra ACIS-S S3 field
of view.  
We corrected these values for Galactic extinction 
as in \citet{schlegel98}.

We then summed over all the tidal features in both samples of galaxies,
to calculate the total r band luminosity 
of the portions of these features covered by the Chandra field of view.
We repeated this process for the disks in each sample, and for
the g band fluxes for the tails/bridges and the disks.
The resultant disk and tail r and g band luminosities $\nu$L$_{\nu}$ 
for each sample are
listed in Table 3.
As can be seen in Table 3, $\sim$13$\%$ of the total g and r
light from the intermediate sample arises from
tidal features.   This is similar to the percentage found for
the \citet{smith10} interacting sample.

Next, for both samples
we compared the 
background-corrected 
number 
of ULX sources above the L$_X$ limit for that sample 
with the r and g luminosities for that sample.
In Table 3, we provide the ratio N(ULX)/L(optical) for each sample,
for the disks and tails separately.
The uncertainties quoted throughout this paper are
statistical uncertainties 
on the number of sources,
calculated
using the approximations for small numbers of events given
by \citet{gehrels86}.  These uncertainties
do not account for 
errors 
in the classification of sources
or in the optical or X-ray luminosities.

These optical luminosities 
should be corrected for internal as well as Galactic extinction.
Using H$\alpha$/H$\beta$ ratios,
for two samples of nearby spiral galaxies
\citet{moustakas06} and 
\citet{moustakas10}
find median E(B $-$ V) values of 0.25
and 0.55, respectively, 
corresponding
to A$_r$ values of 0.7 and 1.5.
For his sample of strongly interacting galaxies, 
\citet{bushouse87}
quotes a typical E(B $-$ V) $\sim$ 0.3,
similar to the above values.
For tidal tails and bridges, only a handful of H$\alpha$/H$\beta$ 
measurements of star forming regions
are available \citep{duc94, duc00,
mirabel92, werk11},
giving a median A$_r$ $\sim$ 0.7, but with considerable scatter.
For the following calculations, we assume that,
on average, galaxian disks 
suffer $\sim$35$\%$ more extinction in the r band than tidal features,
however, 
this is very approximate.

Using these estimates of extinction, 
in Table 3, we provide extinction-corrected values 
of N(ULX)/$\nu$L$_{\nu}$.  
As above, the uncertainties here 
are only statistical uncertainties
on the number of sources, and do not account for uncertainties
in the extinction corrections, which can be substantial.
In Table 3, for both samples, we provide the tidal-to-disk
ratio of the extinction-corrected 
N(ULX)/$\nu$L$_{\nu}$(r) values.   These show a slight 
enhancement
in the tidal features compared to the disks.
However, the uncertainties on the 
N(ULX)/$\nu$L$_{\nu}$ values
are
very large.  
This is shown in Table 3 where we provide the difference between
the tail and disk values, in terms of $\sigma$. 
These tail and disk values agree 
within 2.0$\sigma$.
Furthermore, as noted earlier, there is some uncertainty in the
classification of the ULX candidates as disk vs.\ tidal vs.\ off-galaxy
vs.\ nuclear.
Of our 11 tidal ULXs, four are located in the very extended
and diffuse tidal features of
Arp 215, for which the boundary between `tail' and `off-galaxy'
is ambiguous in some cases.
Furthermore, five of the remaining
tidal ULX candidates are in Arp 270, near
the boundary between `disk' and `tidal', so their classification is
also uncertain.
In addition,
if some of the 
source classified as `nuclear' are actually
`disk' sources, this will slightly decrease the difference between
the disk and tidal sources.  Also, if central AGNs
contribute significantly to the observed optical light, correcting
for that effect will tend to increase the N(ULX)/L(r) ratios of the 
disks compared to the tidal features.

In summary, 
we find a possible slight excess of ULXs associated with tidal
features compared to the inner regions of these interacting galaxies.
Such an enhancement is not unexpected,
if ULX production
is related to star formation.
However, the uncertainties on this result are large, and 
we cannot rule out the possibility that   
tidal features have similar numbers of ULXs relative
to their stellar masses as their parent disks.

\section{Comparison to Spiral Galaxies}

From a sample of 81 nearby galaxies with archival Chandra
data available, \citet{swartz04} concluded that,
on average, spiral galaxies
have larger N(ULX)/L$_B$ ratios
than elliptical galaxies, implying more ULXs per stellar mass.
To compare our sample to their results,
we first need to convert our $\nu$L$_{\nu}$
values
for the SDSS g band to the luminosity in the Johnson B band
L$_B$.  
Using our summation over all the systems 
gives us an average g $-$ r for the disks of 0.51.
For this color, the \citet{windhorst91} magnitude conversion
relation
gives B = g + 0.816.  Combining this with the \citet{zombeck90} conversion
from absolute B magnitude to L$_B$ and a similar equation
for the g band gives
$L_B = 0.486 {\nu}L_{\nu} (g)$.
After converting our luminosities,
in Table 3 we provide N(ULX)/L$_B$ ratios for the disks for our
two samples.

For comparison,
\citet{swartz04}
found 
N(ULX)/L$_B$ = 
7.7 $\pm$ 2.8 $\times$ 10$^{-44}$
ULX/(erg/s) 
for their spiral sample (after correction for a factor
of 3.89 error in their original calculation).
In a more recent study, \citet{swartz11} determined the number
of ULXs per stellar mass for another sample of more nearby galaxies
that was both volume-limited
and magnitude-limited.  This second sample, which
includes mainly spirals and irregular galaxies, gives a slightly lower
ULX-to-optical-light ratio, 
N(ULX)/L$_B$ = 
4.8 $\pm$ 0.5 $\times$ 10$^{-44}$ (erg/s)$^{-1}$.

For comparison to our intermediate sample, we note that
the 
\citet{swartz04} spiral sample has 10 out of 97 ULXs with L$_X$ $\ge$
10$^{40}$ erg/s, giving 7.9 $\pm$ $^{3.4}_{2.5}$ 
$\times$ 10$^{-45}$ (erg/s)$^{-1}$
in this luminosity range.   
In the \citet{swartz11} volume- and magnitude-limit sample,
19 out of 107 ULX candidates have 
L$_X$ $\ge$ 10$^{40}$ erg/s, 
giving a similar value of 
N(ULX)/L$_B$ = 
8.5 $\pm$ $^{2.4}_{1.9}$ $\times$ 10$^{-45}$ (erg/s)$^{-1}$.

In Table 3, we compare our 
N(ULX)/L$_B$ values with the above
values from \citet{swartz04} and \citet{swartz11}.  
There is a significant excess of ULXs (a factor of 2.1 times higher
ratio, a 3.0$\sigma$ effect) in the 
sensitive Arp sample compared to
the second control sample, but less
of an excess compared to the first control sample.
We conclude that, on average, our sample of Arp Atlas galaxies produce
ULXs at a rate 1 $-$ 2 times
than that of spiral galaxies, relative to
the optical luminosity and therefore the stellar mass.

\section{ULX Statistics for Strong Interactions}

The Arp Atlas
contains many strongly interacting galaxies, but it also 
contains some systems that, although peculiar, may not have been strongly
perturbed by a companion in the recent past.
Thus it is possible
that our statistics are being diluted by the presence of 
galaxies in the sample that are not strongly interacting.

To investigate this issue,
from our original set of 45
Arp systems, we select a subset of strongly interacting
systems.   
For each Arp system in the sample, we determined
optical luminosity
ratio(s) for the target galaxy or galaxies compared to the 
nearest neighbor(s), as well as galaxy
separations relative to the diameter of the larger galaxy.
These values were derived from the magnitudes,
sizes, and coordinates available in NED, as 
well as inspection of the images in the Arp Atlas,
SDSS, and Digitized Sky Survey (DSS).
We define a galaxy as strongly interacting if the luminosity
ratio is $\ge$1:4 and the pair separation is $\le$2 diameters
apart.  We also include merger remnants with strong tails
in the strongly interacting category, but exclude 
possible merger remnants without
strong tidal tails (i.e., galaxies with shell-like structures
or with faint tail-like features shorter than one disk diameter).
This criteria gives us 22 Arp systems in
our strongly interacting sample.  These galaxies are identified
in Table 1 by an asterisk.

We then subdivided this set futher into sensitive
and intermediate subsets based on the Chandra sensitivity
criteria used earlier.  The numbers of Arp systems and
individual galaxies in these 
subsets are given in Table 4.  For each of these subsets, 
in Table 4 we 
give the number of ULXs above the 10$^{39}$ and 10$^{40}$
erg/s cutoffs, respectively.
For each subset, we estimated the number of background
sources, determined the respective g and r luminosities,
and calculated N(ULX)/L(optical).  These values are also
given in Table 4.

As in Table 3, in Table 4 we also provide comparisons between
the tidal features and the disks, as well as with
the \citet{swartz04} and \citet{swartz11} values.
The difference between the N(ULX)/L(optical) ratios of
the disks and the tails/bridges is less than 2$\sigma$, thus
we do not find strong evidence of an enhancement of this ratio
in tidal features compared to disks.

However, we do find a $>$2$\sigma$
enhancement of a factor of 2.6 $-$ 4.4
in the N(ULX)/L(opt) in the Arp disks of the strongly
interacting sensitive sample compared to the two control samples.
An enhancement of 
this amount 
would not be surprising if ULX production is related to
star formation.   Interacting galaxies, on average, typically
have mass-normalized star formation rates about two times larger
than more normal galaxies
(e.g., \citealp{bushouse87, bushouse88,
kennicutt87, barton00, barton03, smith07}).  

However, there are other possible explanations.
For example, the optical extinction may be larger in 
interacting galaxies on average than in normal spirals
\citep{bushouse87, bushouse88, smith10b}, perhaps due to interstellar
matter being
driven into the central regions by the interaction.
\citet{bushouse87} found that L$_{FIR}$/L$_{H\alpha}$ for his sample
of strongly interacting galaxies was a factor of $\sim$2.4 $\times$
higher than for his control sample of normal spirals.   If this
is due solely to the optical extinction 
being higher in the interacting vs.\ the
spiral galaxies, this may account for the higher N(ULX)/L(optical)
values in our strongly interacting sample.

\section{Number of ULXs per FIR Luminosity}

Our next goal is to compare the number of ULXs with the current rate
of star formation in these galaxies.
A standard method of estimating the star formation rate
in galaxies is using the far-infrared luminosity (e.g., 
\citealp{kennicutt98}).
The far-infrared luminosities for most nearby galaxies are available
from the IRAS satellite, however, this has relatively low spatial
resolution so generally just a total galaxian flux is available.
If a pair of galaxies is relatively close together, IRAS will provide
a measurement of the combined flux, but not individual fluxes.
In some cases, if the pair of galaxies is widely separated,
individual fluxes for galaxies in pairs is available.

For some of the galaxies in this study, the Chandra field of
view is smaller than the angular size of the galaxy, so only part
of the system is covered by Chandra.   In those cases, we cannot
provide a direct determination of the number of ULXs compared to
the far-infrared luminosity.   Of the 45 Arp systems in our
sample, for 30 the Chandra field covered the whole galaxy
or galaxies in question. 
Arp 169 was not observed by IRAS.
For the remaining 29 systems, we can obtain
both the total far-infrared luminosity and the total number
of ULX sources, so can make a direct comparison.   This sample
of 29 Arp systems we call our IRAS subsample.
These galaxies are identified in Table 1 in the L$_{FIR}$ column.
Of these 29 systems, 16 are in our `strongly interacting' subset.

We then subdivided this IRAS sample into sensitive and intermediate
subsets, based on their Chandra sensitivities as before.
The number of Arp systems in these 
subsets is given in Table 5, along with the number of
individual galaxies covered by the Chandra fields of view,
the mean distance to the galaxies in each sample, and the
sky areas covered.  The number of disk, tail, nuclear, and off-galaxy
X-ray point sources in each sample above their respective 
X-ray luminosity cut-offs are also given in Table 5, along with 
estimated numbers of background contaminants for each subsample
and the number of X-ray sources with optical counterparts.

For each subset,
the total (disk plus tidal)
background-corrected number of non-nuclear ULXs associated
with those systems N(ULX)$_{total}$
(excluding off-galaxy sources) was calculated,
above the X-ray luminosity cut-off
for that sample.
Table 5 also provides the total 
far-infrared luminosity $L_{FIR}$ for each subset, summing
over all of the galaxies in the set. 
The final values of 
N(ULX)$_{total}$/L$_{FIR}$ are given in Table 5.

For comparison, 
the 
\citet{swartz04} sample had 55 galaxies of type S0/a or later,
containing a total of 97 candidate ULXs.   The total L(FIR)
for these galaxies is 1.5 $\times$ 10$^{45}$ erg/s, giving
N(ULX)/L$_{FIR}$ = 6.5 $\pm$ 0.7 $\times$ 10$^{-44}$ (erg/s)$^{-1}$.
For the volume-limited sample of \citet{swartz11},
a similar ratio of 
N(ULX)/L$_{FIR}$ = 5.3 $\pm$ 0.5 $\times$ 10$^{-44}$ (erg/s)$^{-1}$
was found.
These differ from our values for the sensitive sample by less than 50\% (Table 5),
thus on average our Arp galaxies are
producing ULXs at a similar rate relative to their star formation
rates 
as spiral galaxies.
This suggests that the rate of ULX production mainly depends upon
the number of young stars present.  
Environmental effects
such as enhanced cloud collisions, gas compression, 
and shocks in galaxy interactions may increase the rate of star formation
and therefore the number of ULXs, but 
do not provide a large excess of ULXs above what is expected based
on the number of young stars.

In the intermediate sample
at limits of $\ge$10$^{40}$ erg/s,
however,
our Arp systems 
appear
quite deficient in ULXs relative to the far-infrared luminosities,
compared to the spiral samples.
Of the 97 ULX candidates in the \citet{swartz04} sample, 10 have 
L$_X$ $\ge$ 10$^{40}$ erg/s.
This gives 
N(ULX)/L$_{FIR}$ =
6.7 $\pm$ $^{2.9}_{2.1}$ $\times$ 10$^{-45}$ (erg/s)$^{-1}$
above L$_X$ $\ge$ 10$^{40}$ erg/s.
In the more recent \citet{swartz11} sample,
N(ULX)/L$_{FIR}$ =
9.4 $\pm$ 2.2 $\times$ 10$^{-45}$ (erg/s)$^{-1}$
above L$_X$ $\ge$ 10$^{40}$ erg/s.
Comparison with our value in Table 5,
this implies a {\it deficiency} in ULXs in our
intermediate
sample relative 
to the far-infrared luminosities of approximately a factor of 7 $-$ 10
compared to the
comparison samples (a $\sim$2.7 $-$ 3.8$\sigma$ result).

Our intermediate sample contains several merger remnants that
are extremely far-infrared-luminous, luminous enough to be classified
as `luminous
infrared galaxies' (LIRGs) or `ultra-luminous infrared galaxies' (ULIRGs),
including Arp 193, 220, 240, 243, and 299 (see Table 1).
These galaxies
contribute high far-infrared luminosities to the total energy
budgets for the intermediate sample in Table 5, however,
they do not have many 
ULXs above the sensitivity limit of that sample.  
In contrast, the sensitive sample in Table 5 does not contain many
extremely far-infrared luminous galaxies.
The \citet{swartz04} and the \citet{swartz11} samples also do not contain
any LIRGs or ULIRGs.

Our results for the intermediate 
sample implies that there is not a direct proportionality between
the number of very luminous ULXs and the
far-infrared luminosity 
compared to \citet{swartz04, swartz11}.
Instead, 
there is an apparent deficiency 
in high luminosity ULXs for very infrared-luminous
galaxies.
There are a couple of possible explanations for this.   First,
for some of these LIRGs and ULIRGs, an active galactic nucleus may be
contributing significantly to powering the far-infrared luminosity.
If this is the case, the far-infrared may not be directly proportional
to the star formation rate.   Note that some of the
most IR-luminous systems in Table 1
are listed as Seyferts or transition objects.
In some cases, 
mid-infrared spectra
from the Spitzer telescope can provide an indirect estimate of the
fraction 
of the total far-infrared luminosity that is powered by an active nucleus,
via analysis of the mid-infrared spectral features.
For Arp 193, 220, 243, and 299, such an analysis has been conducted, and 
the fraction of the observed far-infrared luminosity
due to an active nucleus has been estimated to be less than 50\%
\citep{veilleux09, petric11, modica12, alonso12}.
Thus AGN domination of interstellar dust heating may not
completely account for the deficiency in ULXs in these galaxies.

Another possibility is that in these
galaxies ULXs are highly obscured in the X-ray, thus they
are not detected above the X-ray flux limits of this survey with our assumptions 
about spectral shape.
Recall that we are not correcting our ULX X-ray luminosities for
internal extinction, which may cause us to underestimate the luminosities in
some cases.
For some starbursts, for example, M82, ULXs have been found
to be very obscured.
Another possibility is that the starbursts that power the
far-infrared luminosity may be too young to have produced ULXs
yet, if ULXs are associated with late O and early B 
stars with main sequence lifetimes of 10 $-$ 20 Myrs.

Thus we conclude that for galaxies with moderate far-infrared
luminosities, as in our sensitive sample,
the number of ULXs is directly related to the far-infrared luminosity.
However, at high L$_{FIR}$, the observed
number of ULXs per L$_{FIR}$ drops off.
This is consistent with earlier studies comparing the galaxy-wide
total X-ray luminosity to FIR luminosity in even more luminous 
LIRGs and ULIRGs, which show a deficiency in the X-ray
at the high FIR end \citep{lehmer10, iwasawa11}.

\section{Local UV/Optical Colors Near
ULXs}

Next, we determined the local UV/optical colors in the vicinity
of the ULXs.
After sky subtraction, we smoothed the SDSS and FUV images to the
spatial resolution of the NUV images (5\farcs6 arcsec).  
At the location of each ULX candidate, 
we then determined
the surface brightness at each UV/optical wavelength, binning
and scaling the SDSS images to match the GALEX pixel size.
At the distances of these galaxies 
(see Table 1), the 5\farcs6 arcsec resolution of these images corresponds
to 0.11 to 6.7 kpc. Thus the size scale of the `local environment'
we are studying in these galaxies varies quite a bit from galaxy
to galaxy.
Our `local' regions are physically much larger
than
the 100 pc scale used in the Swartz et al.\ (2009) study, thus
any correlation of ULXs with
star formation may be more washed out
in our sample compared to that study due to 
larger
contributions from the surrounding stellar populations.

We did this analysis separately for the
four classes of X-ray objects: disk sources, tail/bridge sources,
nuclear sources, and off-galaxy sources.  
We merged the lists
of ULX candidates 
from the sensitive and intermediate samples, removing duplicates.
Histograms of these colors are shown in Figures 2 $-$ 7, 
in the top, second, third, and fourth panels, respectively.
In panels 1, 2, and 4 of Figures 2 $-$ 7, 
we have plotted in red dotted histograms the local colors
for the disk, tail, and off-galaxy sources with SDSS g optical point source
counterparts. 

In Figures 2 $-$ 7, 
these histograms are compared 
with histograms for the global
colors of the sample of normal spiral galaxies studied by
\citet{smith10b}
(sixth panel in each Figure), as well as those
for the main disks of the 
`Spirals, Bridges, and Tails' (SB\&T)
interacting galaxy sample studied by \citet{smith10} (fifth
panel in each Figure).
As seen in Figures 2 $-$ 7 and as previously noted 
\citep{larson78, schombert90, alonso06, smith10b}, 
the optical/UV colors of interacting galaxies
tend to have a larger dispersion than more normal galaxies, likely
due to increased star formation combined with higher extinction.

In general, 
the local UV/optical
colors in the vicinity of the disk ULX candidates 
(top panels in Figures 2 $-$ 7)
have a larger dispersion than
the 
global colors for the 
spiral galaxies (bottom panels),  and in some cases, there is an apparent
shift in the peak of the colors compared to spirals.
In some cases, there is also more dispersion in the local colors 
near the 
disk ULXs compared to the global colors of the SB\&T interacting 
sample. 
Interpreting the distributions of UV/optical colors of the local
vicinities of the ULX candidates is difficult, since they are due to a 
combination of variations in local star formation properties and/or extinction,
combined with possible contamination by UV/optically-bright background
sources.   

The distribution of colors for the off-galaxy sources
may help identify possible background contaminants
in the ULX samples.   However,
as is evident from the fourth panels 
in Figures 2 $-$ 7, the distributions of colors
of the local regions around the 
off-galaxy sources are very uncertain, due to numerous
upper and lower limits. 
Furthermore, a large fraction of the off-galaxy sources
were undetected in both bands of a color, thus are not
included in these plots.

For the off-galaxy sources that are detected in both
bands, the observed colors tend to have a larger spread
compared to the global colors of spirals.
The off-galaxy sources tend to be bluer than the global
colors of galaxies in u $-$ g, but redder in FUV $-$ NUV and i $-$ z.
For g $-$ r and i $-$ z, the off-galaxy sources tend to have 
a large range of colors, both bluer and redder than
the global colors of galaxies.
As can be seen from Figures 2 $-$ 7,
the  
sources with discrete g band counterparts
(marked by the red dotted histograms)
tend to have a larger spread in colors than the sources
without discrete point source optical counterparts,
thus they are more likely to be background objects.

Perhaps the best color 
to use to identify 
background contaminants 
is 
u $-$ g, since that shows the biggest difference in colors between
galaxies as a whole and quasars/AGN.
\citet{richards02} state that using an SDSS color limit of u $-$ g $<$ 0.6
is a good way to find quasars and AGN with z $\le$ 2.2.  
Consistent with this expection,
in Figure 4, most of the detected off-galaxy sources have u $-$ g $<$ 0.6,
while the global colors of galaxies tend to be redder than this limit.
Thus the ULX candidates with optical counterparts that have blue u $-$ g
may be more likely to be background.
However, this is not a definitive cut-off,
since individual star forming regions can also have u $-$ g $<$ 0.6;
for example, in Arp 305, several clumps of star formation 
have 
0.0 $\le$ u $-$ g $\le$ 0.2  \citep{hancock09}.
Thus UV/optical color information alone is insufficient to 
definitively identify background
sources, and followup spectroscopy is necessary to confirm that an
object is a background source.

Even accounting for contamination by background sources, in some cases
the spread in colors for the local regions is larger than
for the global colors.   This is particularly
true for u $-$ g, where
the colors of the local regions around the ULX candidates
are skewed blueward
relative to global colors of both normal spirals and interacting
galaxies (Figure 4).
In Figure 4, 20 out of 58 (34\%) of the disk ULX candidates have
local u $-$ g $<$ 0.6, while almost none of the global colors
of either the spirals or the interacting galaxies are this blue.
This fraction is slightly higher than the expected fraction of 
background contaminants of about 25$\%$ (Table 3).   Thus at
least some of these very blue sources may be true ULXs,
lying within
regions containing younger
stars. 

There is also a shift to the blue 
in the peak of the r $-$ i distribution 
for the disk ULX locations compared to both spirals and interacting 
galaxies (Figure 6).
For the disk ULX locations, 48\% have r $-$ i $<$ 0.2, compared to
14\% of the global colors of the normal spirals, and 22\% of the global
colors of the interacting galaxies.
These differences are unlikely to be due solely to background
contaminants, even assuming that the estimated 25\% interlopers
are all very blue.

For the other colors (NUV $-$ g, g $-$ r, FUV $-$ NUV, and i $-$ z)
there is less difference between the colors of the ULX locations
and the global colors of the interacting galaxies.  The few 
discrepant sources may be background contaminants.

For the 
tidal ULX candidates (2nd panels
in Figures 2 $-$ 7),
the dispersion in colors 
is generally larger than for the disk sources.   
Furthermore, in most colors there is an apparent shift
to the blue
in the mean color of the tidal regions compared to the 
spiral and interacting samples.
This larger spread may be due in part
to contamination by background sources.
In general, a background object 
is likely to have a bigger effect on the optical/UV colors
of a tail or bridge than a higher surface brightness galaxian disk.

However, even accounting for a possible 25\% contamination
rate, the colors of the ULX locations in the tidal features
appear relatively blue. Approximately 38\% of the tidal sources
lie in regions with u $-$ g $<$ 0.6, bluer than most of
the global colors. In g $-$ r, 15\% of the ULX
candidate locations are bluer than g $-$ r = 0.2,
while almost none of the global colors are that blue.
This suggests that the tidal ULXs 
are more likely
to be associated with younger stellar populations. 
Another factor that may cause these colors to be somewhat bluer
is that tidal features tend to have somewhat lower metallicities
than inner disks, which may shift their colors bluewards.

\section{Statistics on Nuclear Sources}

The question of whether active galactic nuclei (AGN)
are more common in interacting
galaxies has been a topic of great interest in the astronomical
community.  Most previous studies have used optical spectroscopy   
to identify active galactic nuclei.
Early studies suggested that there might be an excess of Seyferts in
strongly interacting galaxies \citep{dahari85, keel85}, but other studies 
did not find significant differences in the numbers of Seyferts among strongly
interacting galaxies compared to more isolated systems
\citep{bushouse86, sekiguchi92, donzelli00, casasola04, ellison08}.
Our Chandra study gives us an alternative way to investigate this
question, via X-ray observations that are high enough spatial
resolution to separate out nuclear emission.
Note that not all nuclear X-ray emission may be due to an active
galactic nucleus; 
some of this X-ray radiation
may be from hot gas, ULXs, and/or X-ray binaries
that are associated 
with a nuclear 
starburst.
In selecting our X-ray sources (Section 3.1), we only chose 
objects that were point-like in the Chandra data, eliminating
sources that are extended.   In some cases (for example,
Arp 220) the nuclear X-ray emission is extended (e.g., \citealp{ptak03}),
but may contain a compact point source inside of diffuse hot gas.
Thus our statistics on the nuclear X-ray point sources may be
considered a lower limit, since we may be missing some sources.

The numbers of galactic nuclei in each of our samples
that are detected in the X-ray above
the luminosity limit for that sample are given
in Tables 3, 4, and 5.
These are divided by the total number of individual
galaxies in that sample, to obtain the percent of nuclei
that are detected above that limit.   These percentages are
also given in Tables 3, 4, and 5.
For these calculations we excluded Arp 189, for which the
nucleus was not included in the Chandra S3 field of view.

Of the four sample galaxies classified as Seyfert 1 or 2 in NED,
only one was detected in the X-ray above our flux limits (25\%).
Of the 20 LINER, Sy/LINER, or LINER/HII objects, four were detected above
our flux cut-offs (20\%).
Six of the 20 galactic
nuclei (35\%) optically classified as HII-type were detected in the
X-ray.
In addition, three galaxies without nuclear spectral types in NED
were detected by Chandra above our flux limits (11\%).  
Thus the X-ray detection rates for these
different spectral classes are similar.

To determine whether these percentages are higher than for
more normal galaxies, for comparison
we use the \citet{zhang09} sample of 187 nearby
galaxies.   Their sample is a combination of two samples
of nearby galaxies, both Chandra archive-selected.  The first component
is a volume-limited sample of 
galaxies within 14.5 Mpc, that is also optical and infrared 
magnitude-limited.    The second component is optical magnitude-limited
and angular diameter-limited, and volume-limited to 15 Mpc.
Morphologically, their combined sample
contains 8\% ellipticals, 13\% dwarfs/irregulars,
and the rest spirals.
These percentages are similar to our sample (Section 2).
Their combined sample is 8\% Seyfert and 11\% LINERs, based on optical
spectroscopy.  
The overall percentage of active galaxies in our Arp sample is
somewhat larger, with 6\% Seyfert 1 or 2 and 29\% LINER or transition objects,
however, given the uncertainty in the NED classifications these
percentages are not too inconsistent.
Among our galaxies classified as strongly interacting,
only one is listed as Seyfert 1 or 2 in NED (5\%) and 13 are
LINER or transition objects (59\%).   These percentages are slightly
larger than in our original Arp sample.

Out of the 187 galaxies in the \citet{zhang09}
sample, the nuclei of 29 (15.5 $\pm$ $^{3.5}_{2.9}$\%) were 
detected by Chandra with an X-ray point source luminosity greater than 
10$^{39}$ erg/s.  Above 10$^{40}$ erg/s, 15 (8.0 $\pm$ $^{2.7}_{2.0}$\%) were
detected.
Comparison of these numbers with our values in Table 3
show that the nuclei of our original Arp Atlas
sample of galaxies are detected at 
higher 
rates (about a factor of 2.3 $-$ 3.9 times higher) 
than in this control sample.  These differences are 
1.7$\sigma$ to 3.2$\sigma$.
For the strongly interacting galaxies (Table 4), the percentages
of nuclei that are detected 
are higher than the \citet{zhang09} values by a factor of $\sim$4 $-$ 5,
differences of 
1.3$\sigma$ $-$ 3.2$\sigma$.
Thus X-ray activity appears
enhanced in the nuclei of
interacting galaxies.

The probability that a galaxy is an AGN depends upon its mass and 
therefore its optical
luminosity.   
To roughly account for possible differences in the 
masses and luminosities of the galaxies in the two samples, 
we scale the number
of nuclear X-ray sources with the total
optical luminosity of the sample, as we did earlier for the
disk and tidal sources.
In Tables 3 and 4, for each subset of galaxies
we provide N(nuclear X-ray sources)/L$_B$(total),
the number of nuclear X-ray sources above the X-ray
luminosity limit for that subset, divided by the total (disk plus tail)
g luminosity, converted into B luminosities.
As noted earlier (Section 7), for 15 out of the 45 Arp systems in 
our sample the Chandra fields of view do not cover the full 
extent of the galaxies.  The optical luminosities quoted
in Table 3 and 4 are not the total luminosities, but instead just
the luminosities of the regions covered by the Chandra field of view.
To scale with the number of nuclear sources, in Tables 3 and 4 we roughly correct for 
this difference, to include the entire luminosities of the galaxies
in N(nuclear X-ray sources)/L$_B$(total).

To compare with the \citet{zhang09}
sample, we need to calculate blue luminosities for the \citet{zhang09}
galaxies.   To this end, we extracted blue magnitudes for the \citet{zhang09}
galaxies from NED, preferentially using values from the 
Third Reference Catalogue of Bright Galaxies (RC3; \citealp{devaucouleurs91, 
corwin94}), which were 
available
for
94\% of the galaxies.
We then calculated blue luminosities for these galaxies using the distances
in \citet{zhang09}, along with M$_B$$\sun$ = 5.48 and L$_B$$\sun$ = 1.9 $\times$ 10$^{33}$
erg/s.    This gives a total L$_B$ for the entire \citet{zhang09} sample of
1.97 $\times$ 10$^{45}$ erg/s.   Dividing this number into the number of central
X-ray sources above the two luminosity cut-offs gives 
N(nuclear)/L$_B$$_{total}$ 
values as given in Tables 3 and 4.
The above values for the \citet{zhang09} sample are consistent within
2$\sigma$ uncertainties with the ratios for our samples (Tables 3 and 4).
Thus 
we do not find strong evidence for enhancement of nuclear X-ray activity
in interacting galaxies vs.\ normal galaxies when scaling by the blue 
luminosity.

We also compare the X-ray detection rate 
for the galactic nuclei in our sample with that of more luminous galaxies.
For their sample of 17 LIRGs,
\citet{lehmer10} found that nine nuclei were detected with L$_X$ $\ge$ 10$^{40}$
erg/s.  This percentage of 53 $\pm$ $^{24}_{17}$\% is consistent
with that for our
strongly interacting 
sample. 
Their sample contains more optically-identified
Seyferts than our sample (29\%), but fewer LINER and transition objects (12\%),
giving a similar overall percentage of optically-defined active galaxies. 
For a higher redshift (0.25 $<$ z $<$ 1.05) sample of massive galaxies
(M$_{stellar}$ $>$ 2.5 $\times$ 10$^{10}$ M$_{\sun}$), \citet{silverman11}
determined the fraction of nuclei detected by Chandra with L$_X$
$\ge$ 2 $\times$ 10$^{42}$ erg/s, a higher luminosity cut-off than
we used for our sample.  They found an enhancement of a factor of 
$\sim$2 in their sample of close pairs, compared to a well-matched
control sample.

As noted earlier,
nuclear X-ray emission in the range we 
have studied, with L$_X$ $<$ 10$^{42}$ erg/s,
is not necessarily from an AGN;
a nuclear starburst may have an X-ray-bright nucleus due to
hot gas,
supernovae, ULXs, and/or 
multiple high mass X-ray binaries associated with the young stellar population.
For example, the starburst nucleus of NGC 7714 is unresolved
with Chandra at a distance of 37 Mpc, and has a 0.3 $-$ 8 keV
luminosity of 4.4 $\times$ 10$^{40}$ erg/s \citep{smith05}.
Since interactions may enhance nuclear star formation (e.g., 
\citealp{bushouse87}), 
the star formation
contribution to the X-ray flux from the nucleus
may be stronger in our sample than in the \citet{zhang09} sample.
It is also important to keep in mind that
our sample of galaxies
is significantly more distant than the \citet{zhang09} sample,
thus in some cases circumnuclear star formation 
may
contribute to our quoted 
nuclear X-ray luminosities.    
Possible differences in the morphological types of the galaxies in our sample and those
in the 
\citet{zhang09} sample may also contribute to differences in nuclear
detection rates, as early-type galaxies are more likely to have X-ray bright
nuclei than late-type \citep{zhang09}.
In addition, our Arp sample and/or the \citet{zhang09} sample may be
biased towards AGN since they are archive-selected, depending upon the
original selection criteria for the original proposals.

\section{Summary}

We have identified 71 candidate ULXs
in a set of 
45 peculiar galaxy systems selected
from the Arp Atlas to have both Chandra X-ray and SDSS optical
images.
We find a slight enhancement in 
the number of ULX sources normalized to
the blue luminosity (a factor of $\sim$ two higher)
compared to earlier
studies of spiral galaxies.
This may be due to slightly enhanced star formation in these
galaxies on average, if ULXs are associated with a young
stellar population, or alternatively, to larger average optical extinctions
in these galaxies.
In the tidal features, we find an excess of ULX candidates per optical
luminosity compared to the disks, though ambiguities in classification
of these sources makes this result uncertain.

We also compared the number of ULXs with the far-infrared luminosities,
which are an approximate measure of the star formation rate.
For the Arp systems with the most sensitive Chandra observations,
N(ULX)/L$_{FIR}$ is consistent with that of spirals.   
This suggests that the ULX production rate in these systems
is proportional to the number of young stars.
However, when we extend our sample to include more distant
Arp systems with higher far-infrared luminosities but
less sensitive Chandra observations,
we find a deficiency of high luminosity ULXs compared
to the total far-infrared luminosity.
In some of these infrared-luminous galaxies, an active
galactic nucleus may contribute significantly to dust heating,
lowering the N(ULX)/L$_{FIR}$ ratio.  Alternatively, ULXs
may be more obscured in the X-ray in these extreme systems, most of
which are merger remnants or very close pairs.

In comparing 
the UV/optical colors of the local vicinities of the ULX candidates
with those of interacting galaxies as a whole,
we do not find strong differences, with the exception of 
the u $-$ g and r $-$ i colors.
In these colors, 
there appears to be an excess of blue colors in the vicinity
of the ULX candidates compared to the global colors,
that appears too large to be accounted for solely by background
contamination.   The most likely explanation is that ULXs tend to 
lie in regions with relatively young stellar populations.

We also determined the number of galaxies
with L$_X$(nuclear) greater than our sample luminosity limits,
and compared with a control sample of nearby galaxies
from \citet{zhang09}.
We found that a higher percentage of the nuclei were detected
in our sample, particularly in the strongly interacting
galaxies where $\sim$4 $-$ 5 $\times$ as many galaxies
were  detected
than in the \citet{zhang09} sample.
However, when we scale the number of X-ray detections
with the blue luminosities of the galaxies we do not see a strong
enhancement in our sample compared to the control sample.


%% Included in this acknowledgments section are examples of the
%% AASTeX hypertext markup commands. Use \url without the optional [HREF]
%% argument when you want to print the url directly in the text. Otherwise,
%% use either \url or \anchor, with the HREF as the first argument and the
%% text to be printed in the second.

\acknowledgments

We thank the anonymous referee for helpful suggestions that greatly
improved this paper.
We thank Qiongge Li and Nic Willis for help with downloading 
data,
and Mark Hancock and Mark Giroux for helpful discussions.
This work was funded by NASA Chandra grant AR9-0010A.
D.A.S. is supported in part by NASA Chandra
grants GO9-0098X and GO0-11099A.
This work has made use of the NASA Chandra archives as well as
the Chandra Source Catalog (CSC), provided by the Chandra
X-ray Center (CXC) as part of the Chandra data archive.
This research has made use of the NASA/IPAC Extragalactic Database (NED) 
and the NASA/ IPAC Infrared Science Archive, 
which are operated by the Jet Propulsion Laboratory, 
California Institute of Technology, under contract with the National 
Aeronautics and Space Administration.

\clearpage

\begin{figure}
\plotone{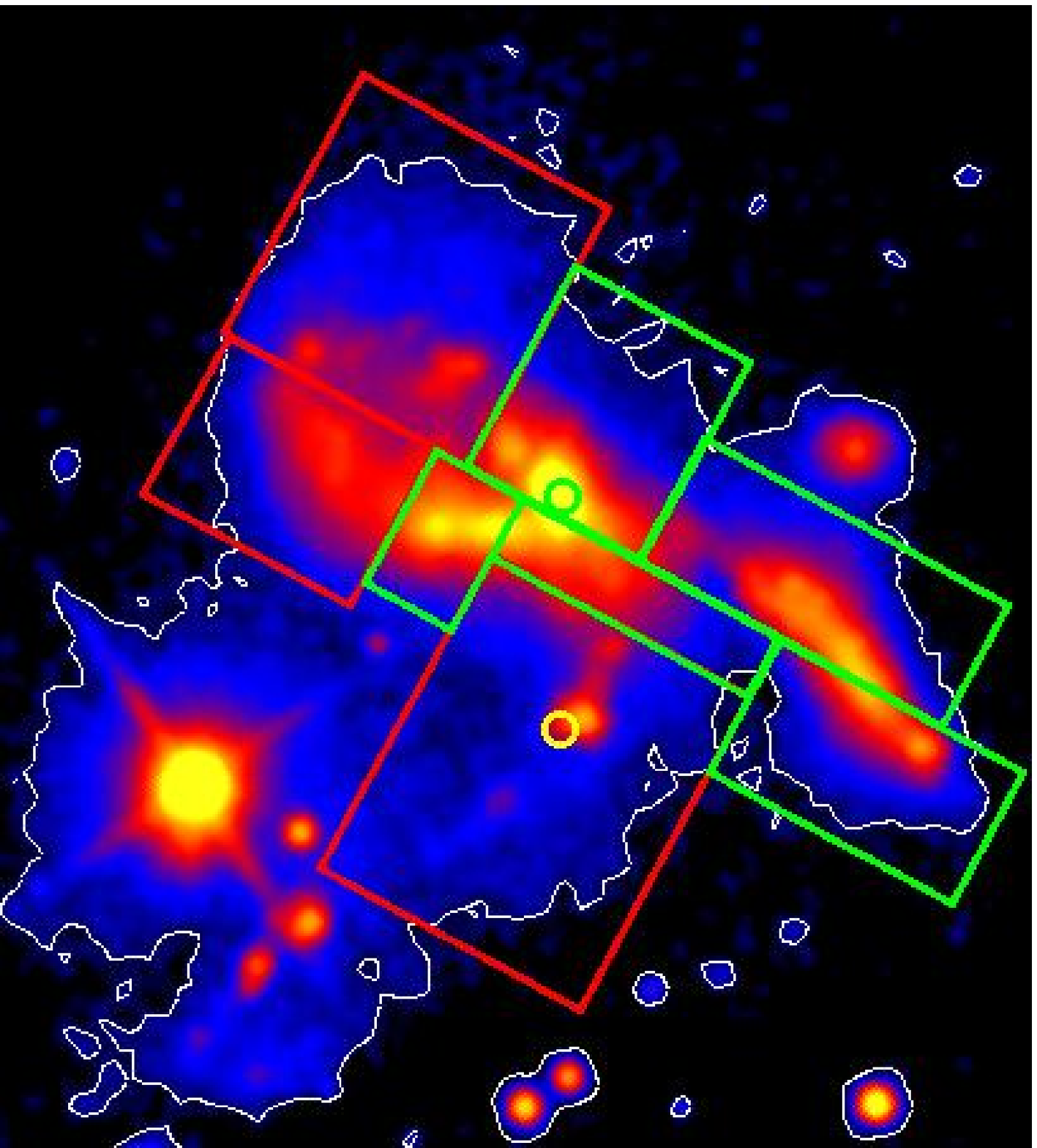}
\caption{
  \small 
The smoothed g-band SDSS image of Arp 259, with ULX candidates marked.
North is up and east to the left.
The source marked in green is a disk ULX candidate, 
while the yellow circle marks the tidal source.
The circles have 2$''$ radii.
The rectangular regions selected as disk are outlined in green,
while the tail areas included are marked in red.   The white contours
mark 2.5 counts about the sky level, which we are using as our dividing
point between `sky' and `galaxy'.
According to \citet{amram07}, this system consists of three 
interacting galaxies.
}
\end{figure}

\begin{figure}
\plotone{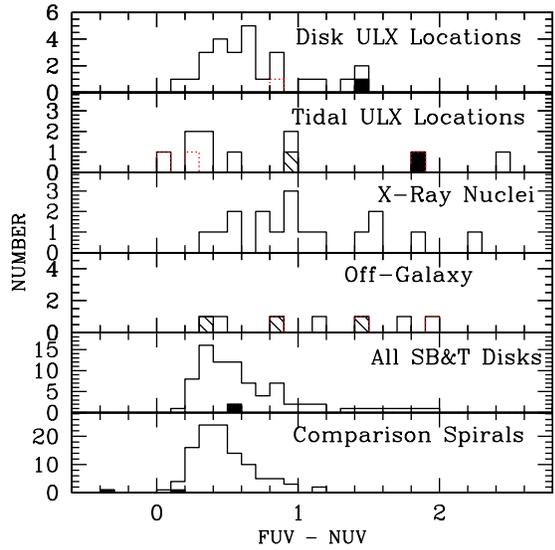}
\caption{
  \small 
Histogram of the FUV $-$ NUV colors for the ULX locations in
the disks (top panel) and the tidal features (second panel),
along with that for the X-ray bright galactic nuclei in
the sample (third panel), and the off-galaxy locations (fourth panel).
These are compared with global disk colors for the SB\&T 
interacting galaxy sample in the fifth panel (from \citealp{smith10})
and normal spirals (from \citealp{smith10b}).
The filled regions are lower limits and the hatched regions are
upper limits.  
The red dotted histograms mark the sources with point
source counterparts on the SDSS g image.
}
\end{figure}

\begin{figure}
\plotone{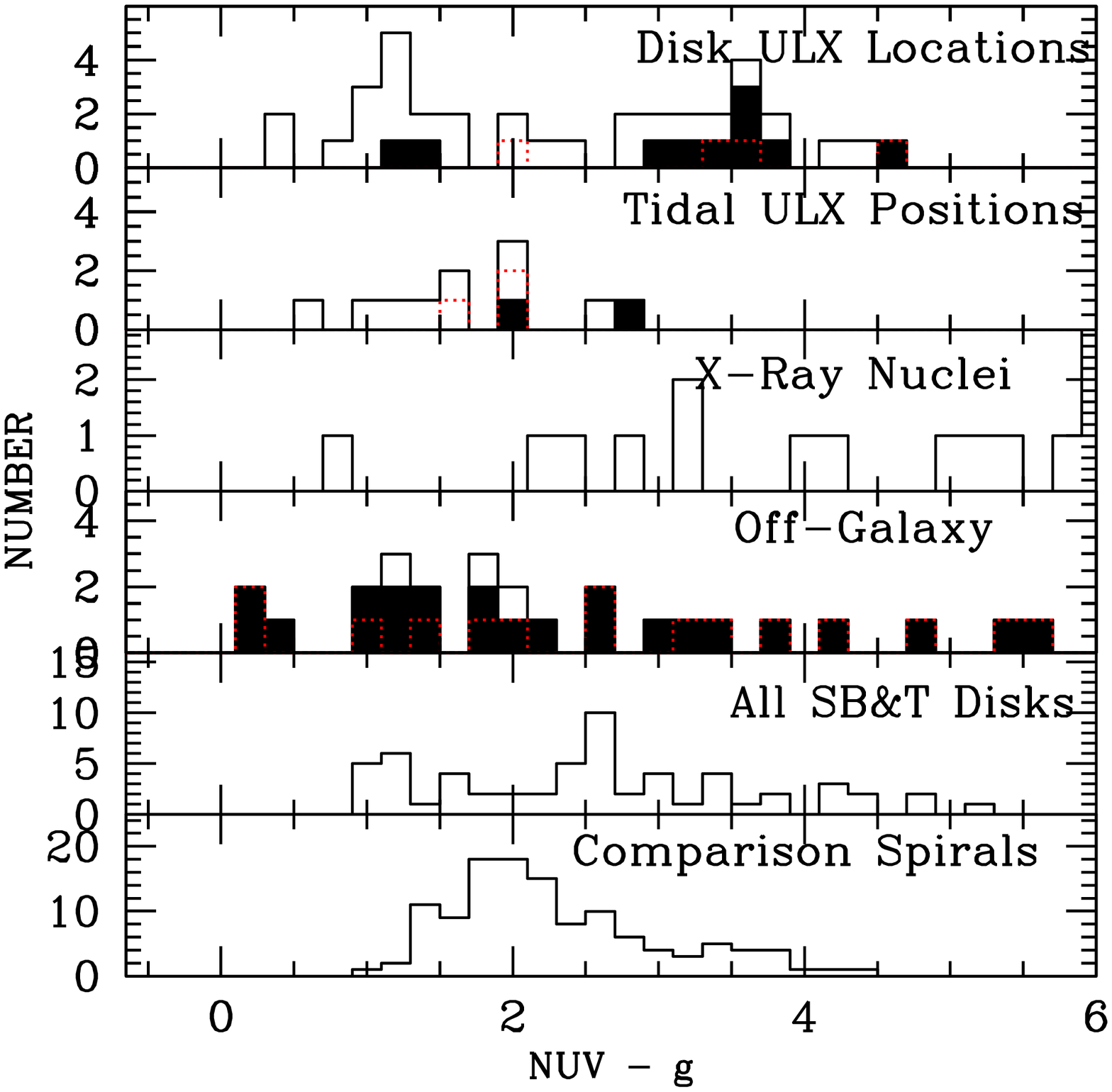}
\caption{
  \small 
Histogram of NUV $-$ g colors as in Figure 2.
The filled regions are lower limits.
The red dotted histograms mark the sources with point
source counterparts on the SDSS g image.
}
\end{figure}

\begin{figure}
\plotone{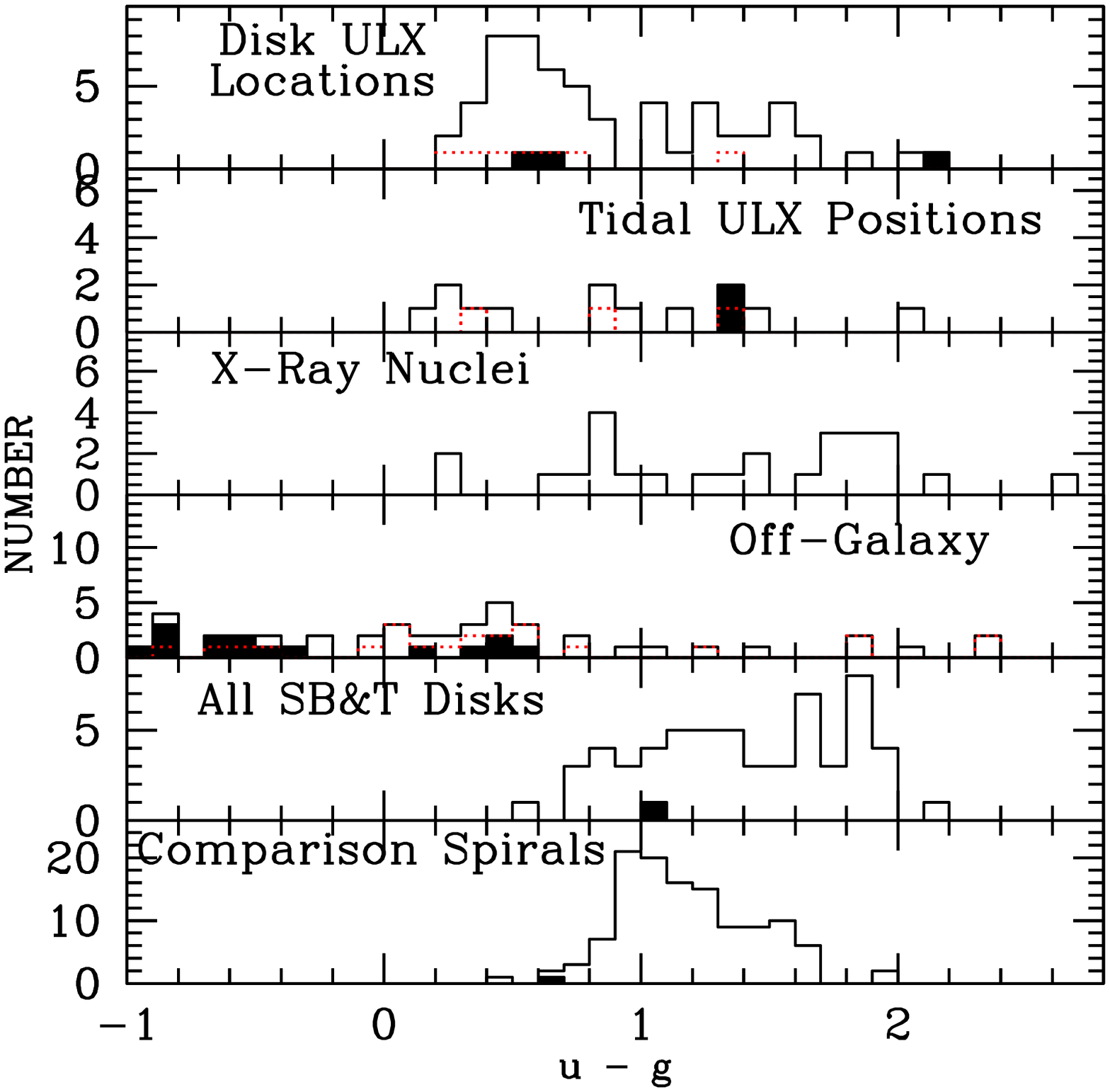}
\caption{
  \small 
Histogram of u $-$ g colors as in Figure 2.
The filled regions are lower limits.
The red dotted histograms mark the sources with point
source counterparts on the SDSS g image.
}
\end{figure}

\begin{figure}
\plotone{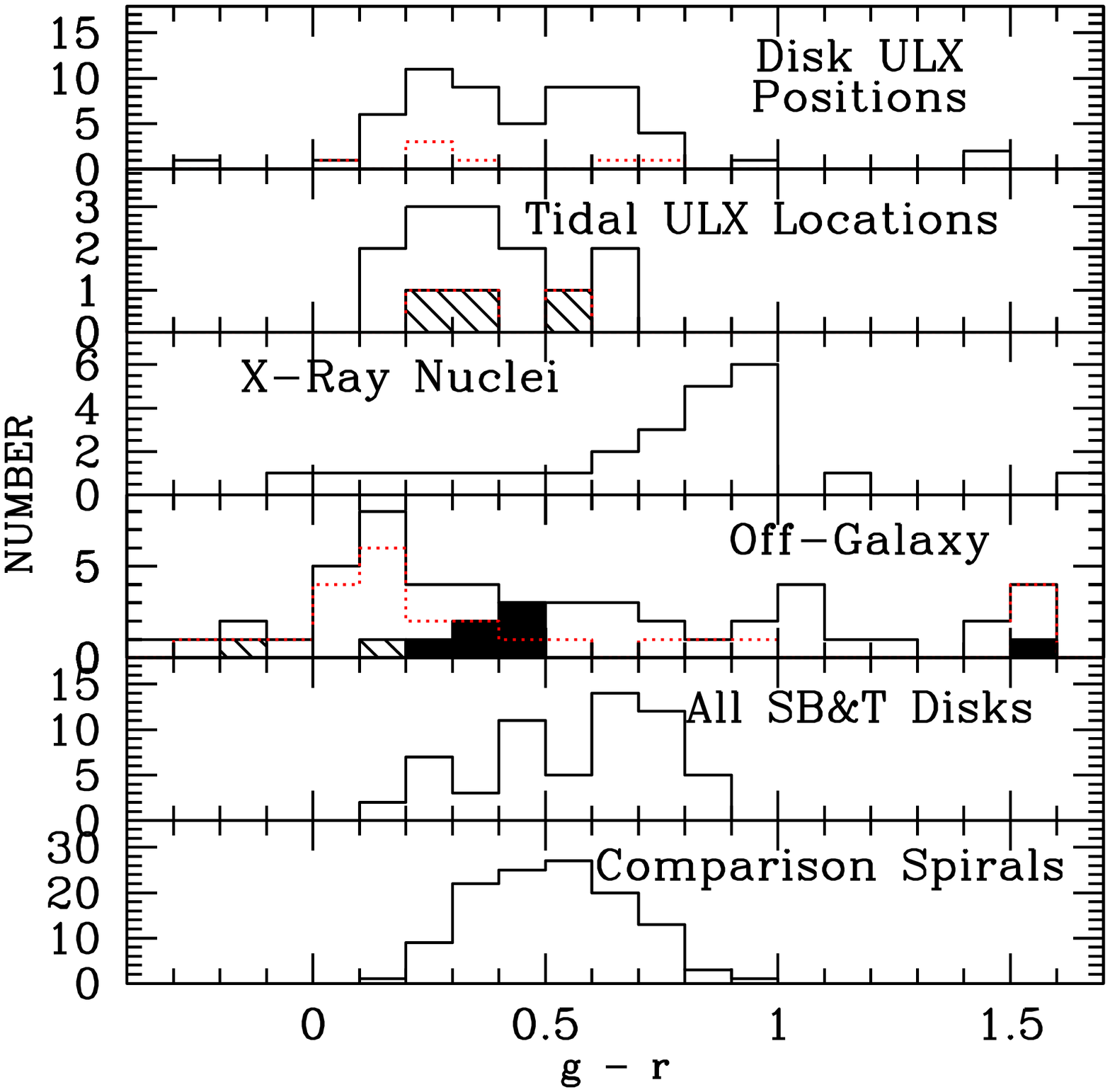}
\caption{
  \small 
Histogram of g $-$ r colors as in Figure 2.
The filled regions are lower limits and the hatched regions are
upper limits.  
The red dotted histograms mark the sources with point
source counterparts on the SDSS g image.
}
\end{figure}

\begin{figure}
\plotone{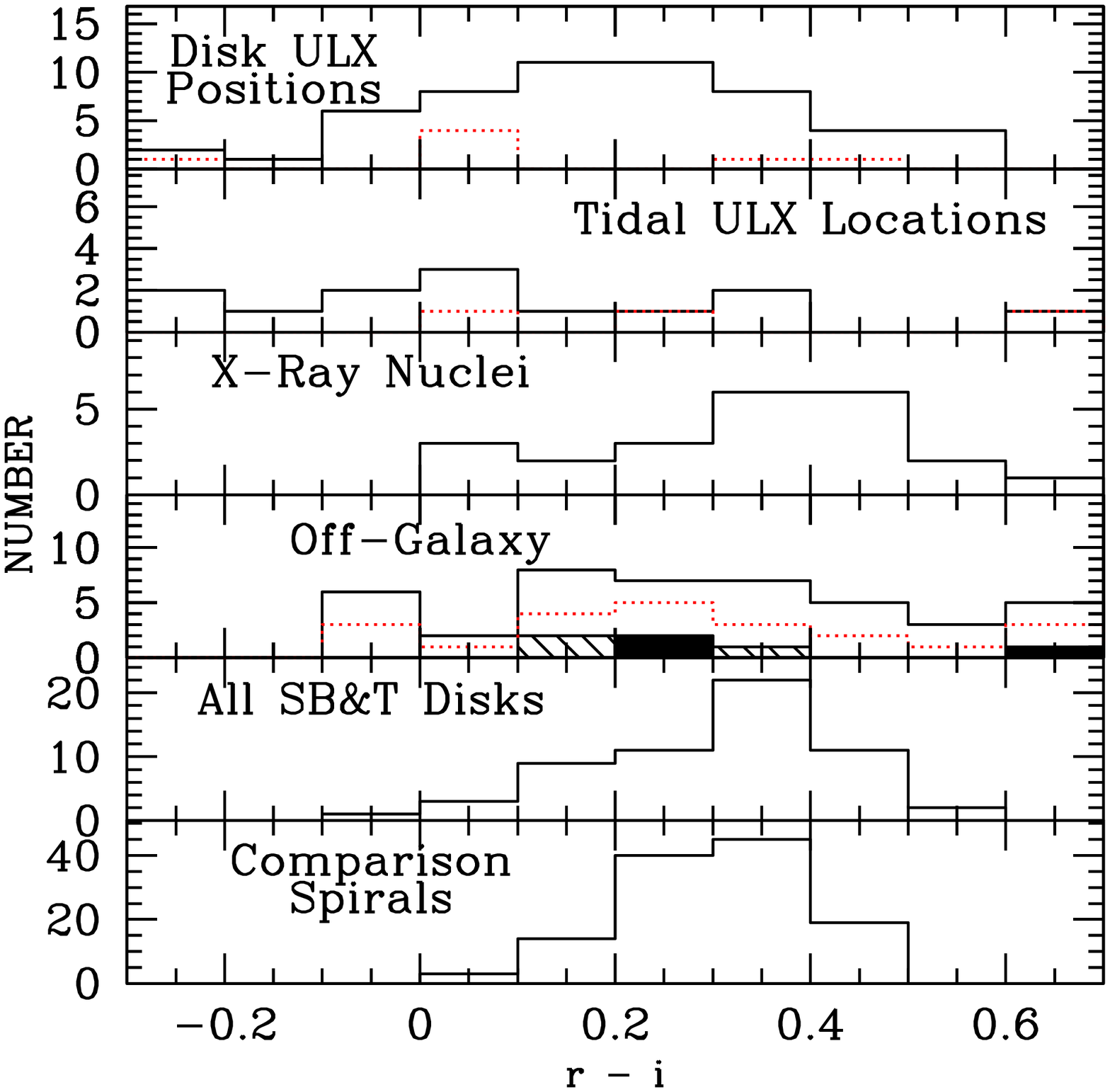}
\caption{
  \small 
Histogram of r $-$ i colors as in Figure 2.
The filled regions are lower limits and the hatched regions are
upper limits.  
The red dotted histograms mark the sources with point
source counterparts on the SDSS g image.
}
\end{figure}

\begin{figure}
\plotone{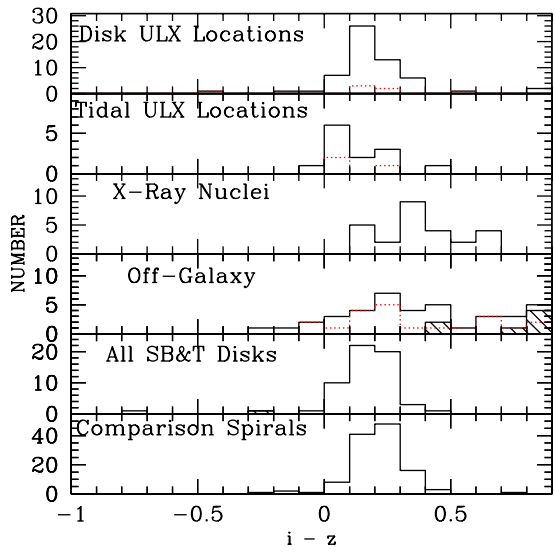}
\caption{
  \small 
Histogram of i $-$ z colors as in Figure 2.
The hatched regions are
upper limits.  
The red dotted histograms mark the sources with point
source counterparts on the SDSS g image.
}
\end{figure}

\clearpage

\begin{deluxetable}{rccrrrrrrrccccccc}
\rotate
\tabletypesize{\scriptsize}
\setlength{\tabcolsep}{0.05in}
\def\et#1#2#3{${#1}^{+#2}_{-#3}$}
\tablewidth{0pt}
\tablecaption{Interacting Galaxy Sample for ULX Study and Available Chandra Data}
\tablehead{
\multicolumn{1}{c}{Arp} &
\multicolumn{1}{c}{Observed} &
\multicolumn{1}{c}{Nuclear} &
\multicolumn{3}{c}{R.A.} &
\multicolumn{3}{c}{Dec.} &
\multicolumn{1}{c}{D$^{\dagger}$}&
\multicolumn{1}{c}{Diameter$^{\ddagger}$} &
\multicolumn{1}{c}{log$^{\amalg}$} &
\multicolumn{1}{c}{Chandra} &
\multicolumn{1}{c}{L$_X$}&
\multicolumn{1}{c}{Chandra}&
\multicolumn{1}{c}{Chandra}&
\multicolumn{1}{c}{GALEX}
\\
\multicolumn{1}{c}{Name} &
\multicolumn{1}{c}{Galaxy} &
\multicolumn{1}{c}{Spectral} &
\multicolumn{3}{c}{(J2000)} &
\multicolumn{3}{c}{(J2000)} &
\multicolumn{1}{c}{(Mpc)} &
\multicolumn{1}{c}{(arcmin)} &
\multicolumn{1}{c}{L$_{FIR}$} &
\multicolumn{1}{c}{Exposure} &
\multicolumn{1}{c}{Limit}&
\multicolumn{1}{c}{Array}&
\multicolumn{1}{c}{Dataset}&
\multicolumn{1}{c}{FUV/NUV}
\\
\multicolumn{1}{c}{} &
\multicolumn{1}{c}{Name(s)} &
\multicolumn{1}{c}{Type} &
\multicolumn{3}{c}{} &
\multicolumn{3}{c}{} &
\multicolumn{1}{c}{} &
\multicolumn{1}{c}{} &
\multicolumn{1}{c}{(L$_{\sun}$)} &
\multicolumn{1}{c}{Time} &
\multicolumn{1}{c}{(log}&
\multicolumn{1}{c}{}&
\multicolumn{1}{c}{}&
\multicolumn{1}{c}{Exposure}
\\ 
\multicolumn{1}{c}{} &
\multicolumn{1}{c}{} &
\multicolumn{1}{c}{} &
\multicolumn{3}{c}{} &
\multicolumn{3}{c}{} &
\multicolumn{1}{c}{} &
\multicolumn{1}{c}{} &
\multicolumn{1}{c}{} &
\multicolumn{1}{c}{(ksec)} &
\multicolumn{1}{c}{$[{\rm erg~s^{-1}}]$}&
\multicolumn{1}{c}{}&
\multicolumn{1}{c}{}&
\multicolumn{1}{c}{Times}
\\ 
\multicolumn{1}{c}{} &
\multicolumn{1}{c}{} &
\multicolumn{1}{c}{} &
\multicolumn{3}{c}{} &
\multicolumn{3}{c}{} &
\multicolumn{1}{c}{} &
\multicolumn{1}{c}{} &
\multicolumn{1}{c}{} &
\multicolumn{1}{c}{} &
\multicolumn{1}{c}{}&
\multicolumn{1}{c}{}&
\multicolumn{1}{c}{}&
\multicolumn{1}{c}{(sec)}
\\ 
}
\startdata
Arp 16   &   M66   &   LINER/Sy2   &   11   &   20   &   15.0   &   +12   &   59   &   30   &   10   &   4 $\times$ 9   &   10.06    &   49.5   &   37.41   &   ACIS-S  & 09548 &  1680,3072 \\
Arp 23   &   NGC 4618   &   HII   &   12   &   41   &   32.8   &   +41   &   9   &   3   &   7.3   &   3 $\times$ 4   &   8.72$^a$   &   54.8   &   37.09   &   ACIS-S  & 09549 &  3242,3242 \\
Arp 24   &   NGC 3445/UGC 6021   &      &   10   &   54   &   40.1   &   +56   &   59   &   11   &   33   &   2 $\times$ 0.8   &   9.6$^a$   &   4.6   &   39.54   &   ACIS-I  & 01686 &  4118,4118 \\
Arp 26   &   M101   &   HII   &   14   &   3   &   12.6   &   +54   &   20   &   57   &   7.2   &   27 $\times$ 29   &   9.46    &   56.2   &   37.07   &   ACIS-S  & 04731 &  1040,1040 \\
Arp 27   &   NGC 3631   &      &   11   &   21   &   2.9   &   +53   &   10   &   10   &   22   &   5 $\times$ 5   &   9.99$^a$   &   89   &   37.84   &   ACIS-S  & 03951 &  1647,1647 \\
Arp 37   &   M77   &   Sy2   &   2   &   42   &   40.7   &   -0   &   0   &   48   &   12   &   6 $\times$ 7   &   10.64    &   72.3   &   37.41   &   ACIS-S  & 00329 &  1616,1616 \\
Arp 76   &   M90   &   LINER/Sy   &   12   &   36   &   49.8   &   +13   &   9   &   46   &   12   &   4 $\times$ 10   &   9.45    &   39.1   &   37.68   &   ACIS-S  & 05911 &  1598,4759 \\
$\ast$ Arp 84   &   NGC 5394/5   &   HII   &   13   &   58   &   35.8   &   +37   &   26   &   20   &   56   &   3.5 $\times$ 1.2   &   10.73    &   15.6   &   39.4   &   ACIS-S  & 10395 &  2820,4286 \\
$\ast$ Arp 91   &   NGC 5953/4   &      &   15   &   34   &   33.7   &   +15   &   11   &   49   &   34   &   1.7   &   10.33$^a$   &   9.9   &   39.18   &   ACIS-S  & 02930 &   \\
$\ast$ Arp 94   &   NGC 3226   &   LINER   &   10   &   23   &   29.8   &   +19   &   52   &   54   &   20   &   6.0   &   9.48    &   2.2   &   39.36   &   ACIS-S  & 01616 &   \\
Arp 104   &   NGC 5216/8   &   Sy1,HII/LINER   &   13   &   32   &   8.9   &   +62   &   44   &   2   &   51   &   5.9 $\times$ 1.9   &   10.49$^a$   &   5.4   &   39.79   &   ACIS-S  & 10568 &   \\
Arp 116   &   NGC 4647/9   &   HII   &   12   &   43   &   36.0   &   +11   &   34   &   2   &   14   &   15   &   9.38    &   38.1   &   37.82   &   ACIS-S  & 00785 &  1658,1658 \\
$\ast$ Arp 120   &   NGC 4438   &   LINER   &   12   &   27   &   43.0   &   +13   &   2   &   38   &   14   &   10 $\times$ 8   &   9.23    &   25   &   38   &   ACIS-S  & 02883 &  6510,6510 \\
Arp 134   &   M49   &   Sy2   &   12   &   29   &   46.7   &   +8   &   0   &   2   &   16   &   8 $\times$ 10   &   $<$7.98    &   39.5   &   37.92   &   ACIS-S  & 00321 &   \\
$\ast$ Arp 147   &   IC 298   &      &   3   &   11   &   18.9   &   +1   &   18   &   53   &   129   &   0.4   &   10.21$^a$   &   24.5   &   39.96   &   ACIS-S  & 11280 &   \\
Arp 155   &   NGC 3656   &   LINER   &   11   &   23   &   38.8   &   +53   &   50   &   10   &   46   &   1.1 $\times$ 0.8   &   9.99$^a$   &   53.8   &   38.7   &   ACIS-S  & 10541 &  1647,1647 \\
$\ast$ Arp 160   &   NGC 4194   &   HII   &   12   &   14   &   9.5   &   +54   &   31   &   37   &   39   &   1 $\times$ 2   &   10.72$^a$   &   35.4   &   38.74   &   ACIS-S  & 07071 &   \\
Arp 162   &   NGC 3414   &   LINER   &   10   &   51   &   16.2   &   +27   &   58   &   30   &   25   &   3 $\times$ 4   &   8.62    &   13.7   &   38.77   &   ACIS-S  & 06779 &   \\
Arp 163   &   NGC 4670   &      &   12   &   45   &   17.2   &   +27   &   7   &   32   &   11   &   1 $\times$ 1   &   8.74$^a$   &   2.6   &   38.77   &   ACIS-S  & 07117 &   \\
$\ast$ Arp 169   &       &   AGN   &   22   &   14   &   46.9   &   +13   &   50   &   24   &   108   &   2.3   &   $-$    &   27   &   39.84   &   ACIS-I  & 05635 &  7103,7104 \\
Arp 189   &   NGC 4651   &   LINER   &   12   &   43   &   42.6   &   +16   &   23   &   36   &   27   &   3 $\times$ 4   &   9.94    &   24.7   &   38.58   &   ACIS-S  & 02096 &  2568,2568 \\
$\ast$ Arp 193   &   IC 883   &      &   13   &   20   &   35.3   &   +34   &   8   &   22   &   103   &   1.3 $\times$ 1.0   &   11.44$^a$   &   14   &   39.98   &   ACIS-S  & 07811 &   \\
Arp 206   &   UGC 5983/NGC 3432   &   LINER/HII   &   10   &   52   &   23.6   &   +36   &   36   &   24   &   13   &   6.6 $\times$ 1.0   &   9.38$^a$   &   1.9   &   39.05   &   ACIS-S  & 07091 &   \\
Arp 214   &   NGC 3718   &   Sy1/LINER   &   11   &   32   &   34.8   &   +53   &   4   &   5   &   17   &   4 $\times$ 8   &   8.74    &   4.9   &   38.87   &   ACIS-S  & 03993 &  1617,1617 \\
$\ast$ Arp 215   &   NGC 2782   &   Sy1/HII   &   9   &   14   &   5.1   &   +40   &   6   &   49   &   37   &   3 $\times$ 4   &   10.31$^a$   &   29.5   &   38.77   &   ACIS-S  & 03014 &  3459,3460 \\
Arp 217   &   NGC 3310   &   HII   &   10   &   38   &   45.8   &   +53   &   30   &   12   &   18   &   2 $\times$ 3   &   10.23$^a$   &   47.1   &   37.94   &   ACIS-S  & 02939 &   \\
$\ast$ Arp 220   &   Arp 220   &   LINER/HII/Sy2   &   15   &   34   &   57.1   &   +23   &   30   &   11   &   83   &   1 $\times$ 2   &   12.03$^a$   &   56.4   &   39.2   &   ACIS-S  & 00869 &   \\
Arp 233   &   UGC 5720   &   HII   &   10   &   32   &   31.9   &   +54   &   24   &   4   &   25   &   0.8 $\times$ 1   &   9.61$^a$   &   19.1   &   38.62   &   ACIS-S  & 09519 &  1680,1680 \\
Arp 235   &   NGC 14   &      &   0   &   8   &   46.4   &   +15   &   48   &   56   &   13   &   2 $\times$ 3   &   8.75$^a$   &   4   &   38.74   &   ACIS-S  & 07127 &   \\
$\ast$ Arp 240   &   NGC 5257/8   &   HII,HII/LINER   &   13   &   39   &   55.2   &   +0   &   50   &   13   &   102   &   3.0 $\times$ 1.1   &   11.29$^a$   &   19.9   &   39.82   &   ACIS-S  & 10565 &   \\
$\ast$ Arp 242   &   NGC 4676A/B   &      &   12   &   47   &   10.1   &   +30   &   43   &   55   &   98   &   2 $\times$ 5   &   10.65$^a$   &   28.5   &   39.63   &   ACIS-S  & 02043 &  1700,1700 \\
$\ast$ Arp 243   &   NGC 2623   &   LINER/Sy2   &   8   &   38   &   24.1   &   +25   &   45   &   17   &   92   &   0.7 $\times$ 2   &   11.46$^a$   &   19.7   &   39.74   &   ACIS-S  & 04059 &  1696,4026 \\
$\ast$ Arp 247   &   IC 2338/9   &      &   8   &   23   &   33.5   &   +21   &   20   &   35   &   77   &   1.5   &   10.31$^a$   &   29.6   &   39.5   &   ACIS-I  & 07937 &   \\
$\ast$ Arp 259   &   NGC 1741   &   HII,HII,HII   &   5   &   1   &   38.3   &   -4   &   15   &   25   &   55   &   1.4   &   10.33$^a$   &   35.5   &   39.04   &   ACIS-S  & 09405 &  1663,7558 \\
$\ast$ Arp 263   &   NGC 3239   &      &   10   &   25   &   4.9   &   +17   &   9   &   49   &   8.1   &   3 $\times$ 5   &   8.65$^a$   &   1.9   &   38.64   &   ACIS-S  & 07094 &  1766,1766 \\
Arp 269   &   NGC 4485/90   &   HII,HII   &   12    &   30    &   33.6    &   +41    &   40    &   17    &   4.6   &   7 $\times$ 5    &   9.22$^a$   &   19.5   &   37.14   &   ACIS-S  & 01579 &  3247,4471 \\
$\ast$ Arp 270   &   NGC 3395/6   &   HII   &   10   &   49   &   52.6   &   +32   &   59   &   13   &   29   &   2 $\times$ 3   &   10.16$^a$   &   19.4   &   38.74   &   ACIS-S  & 02042 &  1487,2660 \\
Arp 281   &   NGC 4627/4631   &   HII   &   12   &   42   &   3.8   &   +32   &   33   &   25   &   6.7   &   12 $\times$ 3   &   9.77    &   59.2   &   36.98   &   ACIS-S  & 00797 &  1680,1680 \\
$\ast$ Arp 283   &   NGC 2798/9   &   HII,HII   &   9   &   17   &   26.9   &   +41   &   59   &   48   &   30   &   3 $\times$ 3   &   10.48$^a$   &   5.1   &   39.35   &   ACIS-S  & 10567 &  2768,4272 \\
$\ast$ Arp 299   &   NGC 3690   &   HII/AGN,HII   &   11   &   28   &   30.4   &   +58   &   34   &   10   &   48   &   10 $\times$ 4   &   11.59$^a$   &   24.2   &   39.16   &   ACIS-I  & 01641 &   \\
$\ast$ Arp 316   &   NGC 3190/3   &   LINER,LINER   &   10   &   18   &   0.5   &   +21   &   48   &   44   &   24   &   16.4   &   9.65    &   7.1   &   39.01   &   ACIS-S  & 11360 &  2078,2078 \\
Arp 317   &   M65   &   LINER   &   11   &   19   &   41.3   &   +13   &   11   &   43   &   7.7   &   43 $\times$ 35   &   8.81$^a$   &   1.7   &   38.64   &   ACIS-S  & 01637 &  1650,1650 \\
$\ast$ Arp 318   &   NGC 835/8   &   Sy2/LINER,HII   &   2   &   9   &   31.3   &   -10   &   9   &   31   &   53   &   6.4   &   10.95    &   13.8   &   39.41   &   ACIS-S  & 10394 &   \\
$\ast$ Arp 319   &   NGC 7317/8/9   &   Sy2   &   22   &   35   &   57.5   &   +33   &   57   &   36   &   89   &   3.2   &   10.15$^a$   &   19.7   &   39.73   &   ACIS-S  & 00789 &  1658,6761 \\
Arp 337   &   M82   &   HII   &   9   &   55   &   52.7   &   +69   &   40   &   46   &   3.9   &   4 $\times$ 11   &   10.47$^a$   &   15.5   &   37.25   &   ACIS-I  & 01302 &  3075,3075 \\
\enddata
{
$^{\dagger}$From the NASA Extragalactic Database (NED), using, as a first preference,
the mean of the distance-independent determinations, and as a second
choice, H$_0$ = 73 km/s/Mpc, with Virgo,
Great Attractor, and SA infall models.
$^{\ddagger}$Total angular extent of Arp system.  From NED, when available;
otherwise, estimated from the SDSS images or the Digital Sky Survey images
available from NED.
$^{\amalg}$Total 42.4 $-$ 122.5 $\mu$m far-infrared luminosity. 
$\ast$In the strongly interacting subset (see Section 7).
$^a$In the IRAS sample (see Section 8).
}
\end{deluxetable}
\clearpage

%\clearpage

\begin{deluxetable}{cccrcc}
%\rotate
\tabletypesize{\scriptsize}
\def\et#1#2#3{${#1}^{+#2}_{-#3}$}
\tablewidth{0pt}
\tablecaption{Table of ULX Candidates and Nuclear X-Ray Sources}
\tablehead{
\multicolumn{1}{c}{Arp} &
\multicolumn{1}{c}{Location} &
\multicolumn{1}{c}{CXO Name$^1$} &
\multicolumn{1}{c}{L$_X$}& 
\multicolumn{1}{c}{Optical} &
\multicolumn{1}{c}{Sample(s)$^2$} 
\\
\multicolumn{1}{c}{System} &
\multicolumn{1}{c}{} &
\multicolumn{1}{c}{(J2000 Coordinates)} &
\multicolumn{1}{c}{(10$^{39}$ erg/s)}& 
\multicolumn{1}{c}{Source?} &
\multicolumn{1}{c}{} 
\\
\multicolumn{1}{c}{} &
\multicolumn{1}{c}{} &
\multicolumn{1}{c}{} &
\multicolumn{1}{c}{(0.5 $-$ 8 keV)}& 
\multicolumn{1}{c}{} &
\multicolumn{1}{c}{} 
\\
}
\startdata
Arp 16  & disk &  CXO  J112018.3+125900  &    1.7  &        &   S  \\
Arp 16  & disk &  CXO  J112020.9+125846  &   11.6  &        &   I     S  \\
Arp 27  & disk &  CXO  J112054.3+531040  &   24.5  &  yes  &   I     S  \\
Arp 27  & disk &  CXO  J112103.0+531013  &    1.2  &        &   S  \\
Arp 27  & disk &  CXO  J112103.4+531048  &    2.1  &        &   S  \\
Arp 27  & disk &  CXO  J112109.5+530927  &    1.1  &        &   S  \\
Arp 27  & disk &  CXO  J112110.2+531012  &    2.1  &        &   S  \\
Arp 27  & disk &  CXO  J112112.7+531045  &    1.6  &  yes  &   S  \\
Arp 37  & nuclear &  CXO  J024240.8-000047  &  254.6  &       &   I     S  \\
Arp 76  & disk &  CXO  J123653.6+131154  &    1.0  &        &   S  \\
Arp 76  & nuclear &  CXO  J123649.8+130946  &    4.0  &       &   S  \\
Arp 84  & nuclear &  NEW  J135837.9+372528  &   11.1  &       &   I    \\
Arp 104  & nuclear &  NEW  J133207.1+624202  &   25.0  &       &   I    \\
Arp 116  & disk &  NEW  J124337.3+113144  &    1.0  &  yes  &   S  \\
Arp 120  & nuclear &  CXO  J122745.6+130032  &    5.2  &       &   S  \\
Arp 134  & disk &  CXO  J122934.5+080032  &    1.2  &        &   S  \\
Arp 134  & disk &  CXO  J123006.6+080202  &    1.9  &        &   S  \\
Arp 147  & disk &  NEW  J031118.0+011902  &   10.0  &        &   I    \\
Arp 147  & nuclear &  NEW  J031119.5+011848  &   23.1  &       &   I    \\
Arp 155  & disk &  NEW  J112335.0+535002  &    1.6  &        &   S  \\
Arp 155  & disk &  NEW  J112336.2+535025  &    1.0  &        &   S  \\
Arp 155  & disk &  NEW  J112337.7+535042  &    1.2  &        &   S  \\
Arp 155  & disk &  NEW  J112341.7+535032  &    1.2  &        &   S  \\
Arp 160  & disk &  CXO  J121406.2+543143  &   12.8  &  yes  &   I     S  \\
Arp 160  & disk &  CXO  J121408.7+543136  &    3.2  &        &   S  \\
Arp 160  & disk &  CXO  J121409.3+543127  &    1.6  &        &   S  \\
Arp 160  & disk &  CXO  J121409.3+543135  &    1.6  &        &   S  \\
Arp 160  & disk &  CXO  J121409.6+543139  &    1.1  &        &   S  \\
Arp 160  & nuclear &  CXO  J121409.6+543136  &   30.5  &       &   I     S  \\
Arp 160  & tidal &  CXO  J121409.7+543215  &   12.0  &        &   I     S  \\
Arp 162  & nuclear &  NEW  J105116.2+275830  &   23.1  &       &   I     S  \\
Arp 163  & nuclear &  CXO  J124517.2+270731  &    1.3  &       &   S  \\
Arp 169  & nuclear &  CXO  J221445.0+135046  &   16.2  &       &   I    \\
Arp 169  & nuclear &  CXO  J221446.9+135027  &   55.9  &       &   I    \\
Arp 193  & nuclear &  CXO  J132035.3+340822  &  154.0  &       &   I    \\
Arp 214  & nuclear &  NEW  J113234.8+530404  &   53.5  &       &   I     S  \\
Arp 215  & disk &  CXO  J091404.2+400738  &    2.5  &        &   S  \\
Arp 215  & disk &  CXO  J091404.4+400748  &    1.2  &        &   S  \\
Arp 215  & disk &  CXO  J091404.8+400650  &    5.2  &        &   S  \\
Arp 215  & disk &  CXO  J091405.1+400642  &    3.5  &        &   S  \\
Arp 215  & disk &  CXO  J091405.3+400628  &    2.9  &        &   S  \\
Arp 215  & disk &  CXO  J091405.4+400652  &    2.2  &        &   S  \\
Arp 215  & disk &  CXO  J091405.5+400648  &    1.4  &        &   S  \\
Arp 215  & disk &  CXO  J091410.9+400715  &    2.5  &        &   S  \\
Arp 215  & nuclear &  CXO  J091405.1+400648  &   41.7  &       &   I     S  \\
Arp 215  & tidal &  CXO  J091408.3+400530  &    3.6  &  yes  &   S  \\
Arp 215  & tidal &  CXO  J091410.1+400619  &    1.0  &        &   S  \\
Arp 215  & tidal &  CXO  J091414.1+400701  &    1.2  &        &   S  \\
Arp 215  & tidal &  CXO  J091416.8+400659  &    2.2  &        &   S  \\
Arp 217  & disk &  NEW  J103843.3+533102  &    5.7  &        &   S  \\
Arp 217  & disk &  NEW  J103844.5+533005  &    1.4  &        &   S  \\
Arp 217  & disk &  NEW  J103844.6+533007  &    1.5  &        &   S  \\
Arp 217  & disk &  NEW  J103844.8+533004  &    3.7  &        &   S  \\
Arp 217  & disk &  NEW  J103846.0+533004  &    6.9  &        &   S  \\
Arp 217  & disk &  NEW  J103846.6+533038  &    1.8  &        &   S  \\
Arp 217  & disk &  NEW  J103846.7+533013  &    2.9  &        &   S  \\
Arp 217  & disk &  NEW  J103847.3+533028  &    5.1  &        &   S  \\
Arp 217  & disk &  NEW  J103850.2+532926  &    4.2  &        &   S  \\
Arp 217  & nuclear &  NEW  J103845.9+533012  &    8.5  &       &   S  \\
Arp 233  & disk &  CXO  J103232.0+542402  &    4.0  &  yes  &   S  \\
Arp 240  & disk &  NEW  J133953.5+005030  &   32.6  &  yes  &   I    \\
Arp 240  & disk &  NEW  J133958.8+005007  &   16.5  &        &   I    \\
Arp 240  & tidal &  NEW  J133955.9+004960  &   24.8  &        &   I    \\
Arp 242  & nuclear &  CXO  J124610.0+304355  &   23.4  &       &   I    \\
Arp 242  & nuclear &  CXO  J124611.2+304321  &   25.5  &       &   I    \\
Arp 243  & nuclear &  CXO  J083824.0+254516  &   92.6  &       &    I    \\
Arp 247  & disk &  CXO  J082333.7+212053  &   11.5  &        &   I    \\
Arp 247  & nuclear &  CXO  J082332.6+212017  &   13.4  &       &   I    \\
Arp 259  & nuclear &  CXO  J050137.7-041529  &   26.8  &       &   I    \\
Arp 259  & tidal &  CXO  J050137.7-041558  &   11.9  &        &   I    \\
Arp 263  & disk &  CXO  J102508.2+170948  &    1.3  &        &   S  \\
Arp 269  & disk &  CXO  J123030.4+414142  &    1.1  &        &   S  \\
Arp 269  & disk &  CXO  J123030.7+413911  &    1.0  &        &   S  \\
Arp 269  & disk &  CXO  J123032.1+413918  &    1.0  &        &   S  \\
Arp 269  & disk &  CXO  J123043.1+413818  &    1.1  &        &   S  \\
Arp 270  & disk &  CXO  J104949.4+325828  &    2.7  &        &   S  \\
Arp 270  & disk &  CXO  J104949.7+325838  &    5.3  &        &   S  \\
Arp 270  & disk &  CXO  J104949.8+325907  &    2.8  &        &   S  \\
Arp 270  & disk &  CXO  J104951.1+325833  &    2.4  &        &   S  \\
Arp 270  & nuclear &  CXO  J104955.0+325927  &    1.3  &       &   S  \\
Arp 270  & tidal &  CXO  J104946.6+325823  &    2.2  &        &   S  \\
Arp 270  & tidal &  CXO  J104947.4+330026  &   12.1  &        &   I     S  \\
Arp 270  & tidal &  CXO  J104952.1+325903  &    4.7  &  yes  &   S  \\
Arp 270  & tidal &  CXO  J104953.2+325910  &    1.5  &        &   S  \\
Arp 270  & tidal &  CXO  J104959.3+325916  &    1.2  &        &   S  \\
Arp 281  & disk &  CXO  J124155.5+323216  &    2.1  &        &   S  \\
Arp 281  & disk &  CXO  J124211.1+323235  &    1.6  &        &   S  \\
Arp 299  & disk &  NEW  J112831.0+583341  &   16.1  &  yes  &   I    \\
Arp 316  & nuclear &  CXO  J101805.6+214956  &   10.4  &       &   I    \\
Arp 317  & disk &  CXO  J111858.4+130530  &    1.2  &        &   S  \\
Arp 318  & nuclear &  CXO  J020924.5-100808  &   45.2  &       &   I    \\
Arp 318  & nuclear &  CXO  J020938.5-100848  &  157.6  &       &    I    \\
Arp 319  & disk &  NEW  J223603.7+335825  &   14.2  &        &   I    \\
Arp 319  & nuclear &  NEW  J223556.7+335756  &   12.3  &       &   I    \\
Arp 319  & nuclear &  NEW  J223603.6+335833  &  296.9  &       &    I    \\
Arp 319  & tidal &  NEW  J223555.7+335739  &   23.1  &  yes  &   I    \\
Arp 337  & disk &  NEW  J095550.2+694047  &    1.4  &        &   S  \\
\enddata
{
$^1$Sources labeled `new' are not in the current
version of the Chandra Source Catalog. $^2${\bf I:} In the intermediate sample.
{\bf S:} In the sensitive sample.
}
\end{deluxetable}
\clearpage

%\clearpage

\begin{deluxetable}{cccc}
%\rotate
\tabletypesize{\scriptsize}
\def\et#1#2#3{${#1}^{+#2}_{-#3}$}
\renewcommand{\arraystretch}{1.0}
\tablewidth{0pt}
\tablecaption{Results on ULX Candidates}
\tablehead{
\multicolumn{1}{c}{} &
\multicolumn{1}{c}{} &
\multicolumn{1}{c}{Sensitive} &
\multicolumn{1}{c}{Intermediate}
\\
\multicolumn{1}{c}{} &
\multicolumn{1}{c}{} &
\multicolumn{1}{c}{Sample} &
\multicolumn{1}{c}{Sample} 
\\
}
\startdata
{\bf General Properties}&L$_X$(limit) (erg/s) &                                                      1 $\times$ 10$^{39}$  &                            1 $\times$ 10$^{40}$                                                                              \\
&Number Arp Systems &                                                                                25  &                                              45                                                                                                \\
&Number Individual Galaxies&                                                                         29  &                                              69                                                                                                \\
&Mean Distance (Mpc) &                                                                               16.6  &                                            37.6                                                                                              \\
&Disk Area (arcmin$^2$) &                                                                             791.8  &                                           901.2                                                                                            \\
&Tail Area (arcmin$^2$) &                                                                              59.8  &                                            91.4                                                                                            \\
&Off-Galaxy Area (arcmin$^2$) &                                                                       846.5  &                                          1911.0                                                                                            \\
\hline   
{\bf Disk Sources}&Number X-Ray Sources $\ge$ L$_X$(limit) &                                         52  &                                               9                                                                                                \\
&Number $\ge$ L$_X$(limit) w/ Optical Counterparts&                                                   5  &                                               4                                                                                                \\
&Percent $\ge$ L$_X$(limit) w/ Optical Counterparts&                                                  9\%  &                                            44\%                                                                                              \\
&Estimaged Number Background $\ge$ L$_X$(limit) &                                                    9.8  &                                             2.8                                                                                               \\
&Percent Background Contaminants $\ge$ L$_X$(limit) &                                                18\%  &                                            31\%                                                                                              \\
&Background-Corrected Number $\ge$ L$_X$(limit) &                                                    42.2  &                                            6.2                                                                                               \\
\hline                                                                                                                                                                                                                                                     
{\bf Tidal Sources}&Number X-Ray Sources $\ge$ L$_X$(limit) &                                        10  &                                               5                                                                                                \\
&Number $\ge$ L$_X$(limit) w/ Optical Counterparts&                                                   2  &                                               1                                                                                                \\
&Percent $\ge$ L$_X$(limit) w/ Optical Counterparts&                                                 20\%  &                                            20\%                                                                                              \\
&Estimated Number Background $\ge$ L$_X$(limit) &                                                    2.3  &                                             2.0                                                                                               \\
&Percent Background Contamination $\ge$ L$_X$(limit) &                                               23\%  &                                            40\%                                                                                              \\
&Background-Corrected Number $\ge$ L$_X$(limit) &                                                    7.7  &                                             3.0                                                                                               \\
\hline                                                                                                                                                                                                                                                     
{\bf Nuclear Sources}&Number X-Ray Sources $\ge$ L$_X$(limit) &                                      10  &                                              21                                                                                                \\
\hline                                                                                                                                                                                                                                                     
{\bf Off-Galaxy Sources}&Number Sources $\ge$ L$_X$(limit) &                                         38  &                                              28                                                                                                \\
&Number $\ge$ L$_X$(limit) w/ Optical Counterparts&                                                  20  &                                              10                                                                                                \\
&Percent $\ge$ L$_X$(limit) w/ Optical Counterparts&                                                 52\%  &                                            35\%                                                                                              \\
\hline                                                                                                                                                                                                                                                     
{\bf Optical Luminosities}&$\nu$L$_{\nu}$(r)(disk)&                                                  11.65   &                                          34.78                                                                                             \\
(10$^{44}$ erg/s)&$\nu$L$_{\nu}$(r)(tail)&                                                           0.59   &                                           5.12                                                                                              \\
&$\nu$L$_{\nu}$(g)(disk)&                                                                            8.76   &                                           25.75                                                                                             \\
&$\nu$L$_{\nu}$(g)(tail)&                                                                            0.52   &                                           4.17                                                                                              \\
\hline                                                                                                                                                                                                                                                     
{\bf Disks vs.\ Tails }&Bkgd-corrected N(ULX)/$\nu$L$_{\nu}$(r)(disk) &                              361.8  $\pm$ $^{70.9}_{61.7  }$ &                  17.7  $\pm$ $^{11.9}_{8.5  }$                                                                     \\
{\bf N(ULX)/L(opt)}&Bkgd-corrected N(ULX)/$\nu$L$_{\nu}$(r)(tail)&                                   1300.7  $\pm$ $^{723.4}_{524.8  }$ &               58.5  $\pm$ $^{66.3}_{42.0  }$                                                                    \\
(10$^{-46}$ (erg/s)$^{-1}$)&Bkgd-corrected N(ULX)/$\nu$L$_{\nu}$(g)(disk)&                           481.3  $\pm$ $^{94.3}_{82.0  }$ &                  23.9  $\pm$ $^{16.0}_{11.4  }$                                                                    \\
&Bkgd-corrected N(ULX)/$\nu$L$_{\nu}$(g)(tail)&                                                      1488.8  $\pm$ $^{828.1}_{600.8  }$ &               72.0  $\pm$ $^{81.6}_{51.7  }$                                                                    \\
&Bkgd- and extinction-corrected N(ULX)/$\nu$L$_{\nu}$(r)(disk) &                                     134.0  $\pm$ $^{26.3}_{22.8  }$ &                    6.6  $\pm$ $^{4.4}_{3.1  }$                                                                     \\
&Bkgd- and extinction-corrected N(ULX)/$\nu$L$_{\nu}$(r)(tail)&                                      660.24  $\pm$ $^{367.2}_{266.4  }$ &               29.71  $\pm$ $^{33.7}_{21.3  }$                                                                   \\
&Ratio Tail/Disk N(ULX)/$\nu$L$_{\nu}$(r) &                                                           4.9  $\pm$ $^{2.9}_{2.2  }$ &                      4.5  $\pm$ $^{5.4}_{5.3  }$                                                                      \\
&Sigma Difference Tail vs. Disk N(ULX)/$\nu$L$_{\nu}$(r) &                                           2.0    &                                           1.1                                                                                               \\
\hline                                                                                                                                                                                                                                                     
{\bf Arp Disks vs.\ Other Samples}&Arp Disks&                                                        990.3  $\pm$ $^{194.0}_{168.7  }$ &                49.2  $\pm$ $^{32.9}_{23.5  }$                                                                    \\
{\bf Bkgd-corrected N(ULX)/L$_B$(disk)}&Swartz et al.\ (2004)&                                       770.0  $\pm$  280.0  &                             79.4  $\pm$ $^{34.0}_{24.6  }$                                                                    \\
(10$^{-46}$ (erg/s)$^{-1}$)&Ratio Arp Disks vs.\ Swartz et al.\ (2004)&                               1.3  $\pm$ $^{0.4}_{0.8  }$ &                      0.6  $\pm$ $^{0.5}_{0.4  }$                                                                      \\
&Sigma Difference Arp Disks vs. Swartz et al.\ (2004)&                                               0.7  &                                             0.7                                                                                               \\
&Swartz et al.\ (2011)&                                                                              480.0  $\pm$  50.0  &                              85.2  $\pm$ $^{24.4}_{19.4  }$                                                                    \\
&Ratio Arp Disks vs.\ Swartz et al.\ (2011)&                                                          2.1  $\pm$ $^{0.4}_{0.4  }$ &                      0.6  $\pm$ $^{0.4}_{0.3  }$                                                                      \\
&Sigma Difference Arp Disks vs. Swartz et al.\ (2011)&                                               2.9  &                                             0.9                                                                                               \\
\hline                                                                                                                                                                                                                                                     
{\bf Nuclear Detections}&Arp Galaxies Percent Nuclei Detected &                                      35.7  $\pm$ $^{15.3}_{11.1  }$ &                   30.9  $\pm$ $^{ 8.3}_{ 6.7  }$                                                                    \\
{\bf Arp Galaxies vs.\ Other Samples}&Zhang et al.\ (2009) Percent Nuclei Detected&                  15.5  $\pm$ $^{ 3.5}_{ 2.9  }$ &                    8.0  $\pm$ $^{ 2.7}_{ 2.0  }$                                                                    \\
&Ratio Percentage Arp Galaxies vs.\ Zhang et al.\ (2009)&                                             2.3  $\pm$ $^{1.1}_{0.9  }$ &                      3.9  $\pm$ $^{1.4}_{1.6  }$                                                                      \\
&Sigma Difference Arp Galaxies vs.\ Zhang et al.\ (2009)&                                            1.7  &                                             3.2                                                                                               \\
(10$^{-46}$ (erg/s)$^{-1}$)&Arp Galaxies N(nuclear X-ray sources)/L$_B$&                             192.8  $\pm$ $^{82.5}_{59.9  }$ &                  125.6  $\pm$ $^{33.9}_{27.2  }$                                                                   \\
                           &Zhang et al.\ (2009) N(nuclear X-ray sources)/L$_B$&                     147.2  $\pm$ $^{32.8}_{27.2  }$ &                  76.1  $\pm$ $^{25.2}_{19.4  }$                                                                    \\
&Ratio N(nuclear)/L$_B$ Arp Galaxies vs.\ Zhang et al.\ (2009)&                                       1.3  $\pm$ $^{0.6}_{0.5  }$ &                      1.6  $\pm$ $^{0.6}_{0.7  }$                                                                      \\
&Sigma Difference Arp Galaxies vs.\ Zhang et al.\ (2009)&                                            0.7                         &                      1.3                                                                                               \\
\enddata
\end{deluxetable}

%\clearpage

\begin{deluxetable}{cccc}
%\rotate
\tabletypesize{\scriptsize}
\def\et#1#2#3{${#1}^{+#2}_{-#3}$}
\renewcommand{\arraystretch}{1.0}
\tablewidth{0pt}
\tablecaption{Results on ULX Candidates: Strongly Interacting Subset of Galaxies}
\tablehead{
\multicolumn{1}{c}{} &
\multicolumn{1}{c}{} &
\multicolumn{1}{c}{Sensitive} &
\multicolumn{1}{c}{Intermediate}
\\
\multicolumn{1}{c}{} &
\multicolumn{1}{c}{} &
\multicolumn{1}{c}{Sample} &
\multicolumn{1}{c}{Sample} 
\\
}
\startdata
{\bf General Properties} &L$_X$(limit) (erg/s) &                                                     1 $\times$ 10$^{39}$  &                            1 $\times$ 10$^{40}$                                                                              \\
&Number Arp Systems &                                                                                5  &                                               22                                                                                                \\
&Number Individual Galaxies &                                                                        6  &                                               40                                                                                                \\
&Mean Distance (Mpc) &                                                                               21.2  &                                            57.7                                                                                              \\
&Disk Area (arcmin$^2$) &                                                                              56.4  &                                           132.9                                                                                            \\
&Tail Area (arcmin$^2$) &                                                                              29.7  &                                            55.9                                                                                            \\
&Off-Galaxy Area (arcmin$^2$) &                                                                       244.0  &                                          1101.2                                                                                            \\
\hline                                                                                                                                                                                                                                                     
{\bf Disk Sources}&Number X-Ray Sources $\ge$ L$_X$(limit) &                                         18  &                                               7                                                                                                \\
&Number $\ge$ L$_X$(limit) w/ Optical Counterparts&                                                   1  &                                               3                                                                                                \\
&Percent $\ge$ L$_X$(limit) w/ Optical Counterparts&                                                  5\%  &                                            42\%                                                                                              \\
&Estimated Number Background $\ge$ L$_X$(limit) &                                                    2.0  &                                             2.4                                                                                               \\
&Percent Background Contaminants $\ge$ L$_X$(limit) &                                                11\%  &                                            34\%                                                                                              \\
&Background-Corrected Number $\ge$ L$_X$(limit) &                                                    16.0  &                                            4.6                                                                                               \\
\hline                                                                                                                                                                                                                                                     
{\bf Tidal Sources}&Number X-Ray Sources $\ge$ L$_X$(limit) &                                        10  &                                               5                                                                                                \\
&Number $\ge$ L$_X$(limit) w/ Optical Counterparts&                                                   2  &                                               1                                                                                                \\
&Percent $\ge$ L$_X$(limit) w/ Optical Counterparts&                                                 20\%  &                                            20\%                                                                                              \\
&Estimated Number Background $\ge$ L$_X$(limit) &                                                    1.4  &                                             1.9                                                                                               \\
&Percent Background Contamination $\ge$ L$_X$(limit) &                                               13\%  &                                            38\%                                                                                              \\
&Background-Corrected Number $\ge$ L$_X$(limit) &                                                    8.6  &                                             3.1                                                                                               \\
\hline                                                                                                                                                                                                                                                     
{\bf Nuclear Sources}&Number X-Ray Sources $\ge$ L$_X$(limit) &                                      4  &                                               17                                                                                                \\
\hline                                                                                                                                                                                                                                                     
{\bf Off-Galaxy Sources}&Number Sources $\ge$ L$_X$(limit) &                                         23  &                                              27                                                                                                \\
&Number $\ge$ L$_X$(limit) w/ Optical Counterparts&                                                  10  &                                               9                                                                                                \\
&Percent $\ge$ L$_X$(limit) w/ Optical Counterparts&                                                 43\%  &                                            33\%                                                                                              \\
&Estimated Number Background $\ge$ L$_X$(limit) &                                                    24.8  &                                            57.0                                                                                              \\
&Percent Observed/Expected Off-Galaxy $\ge$ L$_X$(limit) &                                            92\%  &                                            47\%                                                                                             \\
\hline                                                                                                                                                                                                                                                     
{\bf Optical Luminosities}&$\nu$L$_{\nu}$(r)(disk)&                                                  1.94  &                                            23.46                                                                                             \\
(10$^{44}$ erg/s)&$\nu$L$_{\nu}$(r)(tail)&                                                           0.44  &                                            4.85                                                                                              \\
&$\nu$L$_{\nu}$(g)(disk)&                                                                            1.66  &                                            17.48                                                                                             \\
&$\nu$L$_{\nu}$(g)(tail)&                                                                            0.38  &                                            3.93                                                                                              \\
\hline                                                                                                                                                                                                                                                     
{\bf Disks vs.\ Tails }&Bkgd-corrected N(ULX)/$\nu$L$_{\nu}$(r)(disk) &                              822.5  $\pm$ $^{274.5}_{216.2  }$ &                19.6  $\pm$ $^{16.1}_{11.0  }$                                                                    \\
{\bf N(ULX)/L(opt)}&Bkgd-corrected N(ULX)/$\nu$L$_{\nu}$(r)(tail)&                                   1962.7  $\pm$ $^{971.9}_{705.1  }$ &               63.5  $\pm$ $^{70.1}_{44.4  }$                                                                    \\
(10$^{-46}$ (erg/s)$^{-1}$)&Bkgd-corrected N(ULX)/$\nu$L$_{\nu}$(g)(disk)&                           961.2  $\pm$ $^{320.7}_{252.7  }$ &                26.2  $\pm$ $^{21.6}_{14.7  }$                                                                    \\
&Bkgd-corrected N(ULX)/$\nu$L$_{\nu}$(g)(tail)&                                                      2279.5  $\pm$ $^{1128.7}_{818.9  }$ &              78.4  $\pm$ $^{86.5}_{54.8  }$                                                                    \\
&Bkgd- and extinction-corrected N(ULX)/$\nu$L$_{\nu}$(r)(disk) &                                     304.6  $\pm$ $^{101.7}_{80.1  }$ &                   7.2  $\pm$ $^{6.0}_{4.1  }$                                                                     \\
&Bkgd- and extinction-corrected N(ULX)/$\nu$L$_{\nu}$(r)(tail)&                                      996.30  $\pm$ $^{493.3}_{357.9  }$ &               32.25  $\pm$ $^{35.6}_{22.6  }$                                                                   \\
&Ratio Tail/Disk N(ULX)/$\nu$L$_{\nu}$(r) &                                                           3.3  $\pm$ $^{1.8}_{1.7  }$ &                      4.5  $\pm$ $^{5.3}_{6.5  }$                                                                      \\
&Sigma Difference Tail vs. Disk N(ULX)/$\nu$L$_{\nu}$(r) &                                           1.9  &                                             1.1                                                                                               \\
\hline                                                                                                                                                                                                                                                     
{\bf Arp Disks vs.\ Other Samples}&Arp Disks&                                                        1977.7  $\pm$ $^{659.9}_{519.9  }$ &               54.0  $\pm$ $^{44.5}_{30.3  }$                                                                    \\
{\bf Bkgd-corrected N(ULX)/L$_B$(disk)}&Swartz et al.\ (2004)&                                       770.0  $\pm$  280.0  &                             79.4  $\pm$ $^{34.0}_{24.6  }$                                                                    \\
(10$^{-46}$ (erg/s)$^{-1}$)&Ratio Arp Disks vs.\ Swartz et al.\ (2004)&                               2.6  $\pm$ $^{0.9}_{0.7  }$ &                      0.7  $\pm$ $^{0.6}_{0.5  }$                                                                      \\
&Sigma Difference Arp Disks vs. Swartz et al.\ (2004)&                                               2.3  &                                             0.5                                                                                               \\
&Swartz et al.\ (2011)&                                                                              480.0  $\pm$  50.0  &                              85.2  $\pm$ $^{24.4}_{19.4  }$                                                                    \\
&Ratio Arp Disks vs. Swartz et al.\ (2011)&                                                           4.1  $\pm$ $^{1.4}_{1.1  }$ &                      0.6  $\pm$ $^{0.5}_{0.4  }$                                                                      \\
&Sigma Difference Arp Disks vs. Swartz et al.\ (2011)&                                               2.9  &                                             0.6                                                                                               \\
\hline                                                                                                                                                                                                                                                     
{\bf Nuclear Detections}&Arp Galaxies Percent Nuclei Detected &                                      66.7  $\pm$ $^{33.3}_{38.2  }$ &                   42.5  $\pm$ $^{13.4}_{10.5  }$                                                                    \\
{\bf Arp Galaxies vs.\ Other Samples}&Zhang et al.\ (2009) Percent Nuclei Detected&                  15.5  $\pm$ $^{ 3.5}_{ 2.9  }$ &                    8.0  $\pm$ $^{ 2.7}_{ 2.0  }$                                                                    \\
&Ratio Percentage Arp Galaxies vs.\ Zhang et al.\ (2009)&                                             4.3  $\pm$ $^{2.3}_{2.6  }$ &                      5.3  $\pm$ $^{2.1}_{2.2  }$                                                                      \\
&Sigma Difference Arp Galaxies vs.\ Zhang et al.\ (2009)&                                            1.3  &                                             3.2                                                                                               \\
(10$^{-46}$ (erg/s)$^{-1}$)&Arp Galaxies N(nuclear X-ray sources)/L$_B$&                             350.7  $\pm$ $^{278.7}_{167.4  }$ &                142.1  $\pm$ $^{43.6}_{34.1  }$                                                                   \\
                           &Zhang et al.\ (2009) N(nuclear X-ray sources)/L$_B$&                     147.2  $\pm$ $^{32.8}_{27.2  }$ &                  76.1  $\pm$ $^{25.2}_{19.4  }$                                                                    \\
&Ratio N(nuclear)/L$_B$ Arp Galaxies vs.\ Zhang et al.\ (2009)&                                       2.4  $\pm$ $^{1.9}_{1.3  }$ &                      1.9  $\pm$ $^{0.7}_{0.8  }$                                                                      \\
&Sigma Difference Arp Galaxies vs.\ Zhang et al.\ (2009)&                                            1.2  &                                             1.6                                                                                               \\
\enddata
\end{deluxetable}

\begin{deluxetable}{cccc}
%\rotate
\tabletypesize{\scriptsize}
\def\et#1#2#3{${#1}^{+#2}_{-#3}$}
\tablewidth{0pt}
\renewcommand{\arraystretch}{1.0}
\tablecaption{Results on ULX Candidates: Comparison with IRAS}
\tablehead{
\multicolumn{1}{c}{} &
\multicolumn{1}{c}{} &
\multicolumn{1}{c}{Sensitive} &
\multicolumn{1}{c}{Intermediate}
\\
\multicolumn{1}{c}{} &
\multicolumn{1}{c}{} &
\multicolumn{1}{c}{Sample} &
\multicolumn{1}{c}{Sample} 
\\
}
\startdata
{\bf General Properties}&L$_X$(limit) (erg/s) &                                                      1 $\times$ 10$^{39}$  &                            1 $\times$ 10$^{40}$                                                                              \\
&Number Arp Systems &                                                                                14  &                                              29                                                                                                \\
&Number Individual Galaxies &                                                                        16  &                                              47                                                                                                \\
&Mean Distance (Mpc) &                                                                               18.1  &                                            43.6                                                                                              \\
&Disk Area (arcmin$^2$) &                                                                             158.4  &                                           218.2                                                                                            \\
&Tail Area (arcmin$^2$) &                                                                              43.4  &                                            72.3                                                                                            \\
&Off-Galaxy Area (arcmin$^2$) &                                                                       676.3  &                                          1560.2                                                                                            \\
\hline                                                                                                                                                                                                                                                     
{\bf Disk Sources}&Number X-Ray Sources $\ge$ L$_X$(limit) &                                         44  &                                               8                                                                                                \\
&Number $\ge$ L$_X$(limit) w/ Optical Counterparts&                                                   4  &                                               4                                                                                                \\
&Percent $\ge$ L$_X$(limit) w/ Optical Counterparts&                                                  9\%  &                                            50\%                                                                                              \\
&Estimated Number Background $\ge$ L$_X$(limit) &                                                    3.2  &                                             1.3                                                                                               \\
&Percent Background Contaminants $\ge$ L$_X$(limit) &                                                 7\%  &                                            15\%                                                                                              \\
&Background-Corrected Number $\ge$ L$_X$(limit) &                                                    40.8  &                                            6.7                                                                                               \\
\hline                                                                                                                                                                                                                                                     
{\bf Tidal Sources}&Number X-Ray Sources $\ge$ L$_X$(limit) &                                        10  &                                               5                                                                                                \\
&Number $\ge$ L$_X$(limit) w/ Optical Counterparts&                                                   2  &                                               1                                                                                                \\
&Percent $\ge$ L$_X$(limit) w/ Optical Counterparts&                                                 20\%  &                                            20\%                                                                                              \\
&Estimated Number Background $\ge$ L$_X$(limit) &                                                    1.9  &                                             1.9                                                                                               \\
&Percent Background Contamination $\ge$ L$_X$(limit) &                                               19\%  &                                            38\%                                                                                              \\
&Background-Corrected Number $\ge$ L$_X$(limit) &                                                    8.1  &                                             3.1                                                                                               \\
\hline                                                                                                                                                                                                                                                     
{\bf Nuclear Sources}&Number X-Ray Sources $\ge$ L$_X$(limit) &                                      5  &                                               12                                                                                                \\
\hline                                                                                                                                                                                                                                                     
{\bf Off-Galaxy Sources}&Number Sources $\ge$ L$_X$(limit) &                                         36  &                                              21                                                                                                \\
&Number $\ge$ L$_X$(limit) w/ Optical Counterparts&                                                  19  &                                               7                                                                                                \\
&Percent $\ge$ L$_X$(limit) w/ Optical Counterparts&                                                 52\%  &                                            33\%                                                                                              \\
&Estimated Number Background $\ge$ L$_X$(limit) &                                                    43.9  &                                            44.8                                                                                              \\
&Percent Observed/Expected Off-Galaxy $\ge$ L$_X$(limit) &                                            82\%  &                                            46\%                                                                                             \\
\hline                                                                                                                                                                                                                                                     
{\bf Disk+Tidal}&Arp Galaxies L(FIR) (10$^{44}$ erg/s)&                                              6.39     &                                         102.33                                                                                            \\
(10$^{-47}$ (erg/s)$^{-1}$) &Arp Galaxies N(ULX)/L(FIR) &                                            7649.0  $\pm$ $^{1315.1}_{1146.6  }$ &             95.7  $\pm$ $^{46.0}_{34.7  }$                                                                    \\
 &Swartz et al.\ (2004) N(ULX)/L(FIR)&                                                               6466.7  $\pm$ $^{725.8}_{655.3  }$ &               666.7  $\pm$ $^{285.2}_{206.9  }$                                                                 \\
 & ratio Arp vs.\ Swartz et al.\ (2004) N(ULX)/L(FIR) &                                               1.2  $\pm$ $^{0.2}_{0.2  }$ &                      0.14  $\pm$ $^{0.08}_{0.08  }$                                                                   \\
 & sigma Difference Arp vs.\ Swartz et al.\ (2004) N(ULX)/L(FIR) &                                   0.9  &                                             2.7                                                                                               \\
 &Swartz et al.\ (2011) N(ULX)/L(FIR)&                                                               5300.0  $\pm$  500.0   &                           940.0  $\pm$  220.0                                                                               \\
 & ratio Arp vs.\ Swartz et al.\ (2011) N(ULX)/L(FIR) &                                               1.4  $\pm$ $^{0.2}_{0.2  }$ &                      0.10  $\pm$ $^{0.05}_{0.05  }$                                                                   \\
 & sigma Difference Arp vs.\ Swartz et al.\ (2011) N(ULX)/L(FIR) &                                   2.0  &                                             3.8                                                                                               \\
\enddata
\end{deluxetable}

%% The following command ends your manuscript. LaTeX will ignore any text

\end{document}